\documentclass[10pt,sort&compress]{article}
\usepackage{graphicx,rotating,multirow}
\usepackage{bm,multirow}
\usepackage{amsmath,amssymb,color,mathrsfs}

\usepackage{footnote}
\usepackage{citesort}
\usepackage{cuted}

\newcommand{\AJP}{ {\em Am. J. Phys. }}
\newcommand{\AnM}{ {\em Annals Math. }}
\newcommand{\APNY}{ {\em Ann. Phys. (N.Y.)} }
\newcommand{\APB}{ {\em Ann. Phys. (Berlin) }}
\newcommand{\CMP}{ {\em Commun. Math. Phys. }}
\newcommand{\CRA}{ {\em C. R. Acad. Sci. Ser. A }}
\newcommand{\EPJB}{ {\em Eur. Phys. J. B }}
\newcommand{\EPL}{ {\em Europhys. Lett. }}
\newcommand{\EJP}{{\em Eur. J. Phys.} }
\newcommand{\FP}{ {\em Found. Phys. }}
\newcommand{\IJQC}{ {\em Int. J. Quantum Chem. }}
\newcommand{\JMC}{ {\em J. Math. Chem. }}
\newcommand{\JMP}{ {\em J. Math. Phys.} }
\newcommand{\jpa}{ {\em J. Phys. A} }
\newcommand{\JPA}{ {\em J. Phys. A} }
\newcommand{\jpg}{ {\em J. Phys. G} }
\newcommand{\JSTAT}{ {\em J. Stat. Mech. }}
\newcommand{\JSP}{ {\em J. Stat. Phys. }}
\newcommand{\JRLR}{ {\em J. Russ. Laser Res. }}
\newcommand{\LMP}{ {\em Lett. Math. Phys. }}
\newcommand{\PA}{ { \em Physica A }}
\newcommand{\PLA}{ {\em Phys. Lett. A }}
\newcommand{\PNAS}{ {\em P. Natl. Acad. Sci. USA }}
\newcommand{\PR}{ {\em Phys. Rev. }}
\newcommand{\PRA}{ {\em Phys. Rev. A }}
\newcommand{\PRB}{ {\em Phys. Rev. B }}
\newcommand{\PRC}{ {\em Phys. Rev. C }}
\newcommand{\PRD}{ {\em Phys. Rev. D }}
\newcommand{\PRE}{ {\em Phys. Rev. E }}
\newcommand{\PRL}{ {\em Phys. Rev. Lett.} }

\begin{document}
\title{R\'{e}nyi and Tsallis entropies: three analytic examples}
\author{O Olendski\footnote{Department of Applied Physics and Astronomy, University of Sharjah, P.O. Box 27272, Sharjah, United Arab Emirates;E-mail: oolendski@sharjah.ac.ae}}

\maketitle

\begin{abstract}
A comparative study of one-dimensional quantum structures which allow analytic expressions for the position and momentum R\'{e}nyi $R(\alpha)$ and Tsallis $T(\alpha)$ entropies, focuses on extracting the most characteristic physical features of these one-parameter functionals. Consideration of the harmonic oscillator reconfirms a special status of the Gaussian distribution: at any parameter $\alpha$ it converts into the equality both R\'{e}nyi and Tsallis uncertainty relations removing for the latter an additional requirement $1/2\leq\alpha\leq1$ that is a necessary condition for all other geometries. It is shown that the lowest limit of the semi-infinite range of the dimensionless parameter $\alpha$ where \emph{momentum} components exist strongly depends on the \emph{position} potential and/or boundary condition for the \emph{position} wave function. Asymptotic limits reveal that in either space the entropies $R(\alpha)$ and $T(\alpha)$ approach their Shannon counterpart, $\alpha=1$, along different paths. Similarities and differences between the two entropies and their uncertainty relations are exemplified. Some unsolved problems are also indicated.
\end{abstract}
\vskip.7in

\noindent

\section{Introduction}\label{Sec_Intro}
To describe how much we know about a location and motion of a nano object, quantum-information theory operates with some functionals of the position $\rho_n(\bf r)$ and momentum\footnote{To keep the results in the most symmetric form, throughout the whole research instead of momentum $\bf p$, if not specified otherwise, we will actually operate with the corresponding wave vector ${\bf k}\equiv{\bf p}/\hbar$.} $\gamma_n(\bf k)$ densities that are squared amplitudes of the corresponding one-particle wavefunctions $\Psi_n({\bf r})$ and $\Phi_n({\bf k})$:
\begin{subequations}\label{Densities1}
\begin{eqnarray}
\label{DensityX1}
\rho_n({\bf r})&=\left|\Psi_n({\bf r})\right|^2\\
\label{DensityP1}
\gamma_n({\bf k})&=\left|\Phi_n({\bf k})\right|^2,
\end{eqnarray}
\end{subequations}
where a discrete index $n$ counts all possible bound quantum states. In general $l$-dimensional space, $\Psi_n({\bf r})$ and $\Phi_n({\bf k})$ are related through the Fourier transformation:
\begin{subequations}\label{Fourier1}
\begin{eqnarray}\label{Fourier1_1}
\Phi_n({\bf k})&=\frac{1}{(2\pi)^{l/2}}\int_{\mathbb{R}^l}\Psi_n({\bf r})e^{-i{\bf kr}}d{\bf r},\\
\label{Fourier1_2}
\Psi_n({\bf r})&=\frac{1}{(2\pi)^{l/2}}\int_{\mathbb{R}^l}\Phi_n({\bf k})e^{i{\bf rk}}d{\bf k},
\end{eqnarray}
\end{subequations}
with integrations carried out over the whole available region. Both of them satisfy orthonormality conditions:
\begin{equation}\label{OrthoNormality1}
\int_{\mathbb{R}^l}\Psi_{n'}^\ast({\bf r})\Psi_n({\bf r})d{\bf r}=\int_{\mathbb{R}^l}\Phi_{n'}^\ast({\bf k})\Phi_n({\bf k})d{\bf k}=\delta_{nn'},
\end{equation}
where $\delta_{nn'}=\left\{\begin{array}{cc}
1,&n=n'\\
0,&n\neq n'
\end{array}\right.$ is a Kronecker delta, $n,n'=1,2,\ldots$.
Position wave functions $\Psi_n({\bf r})$ and associated eigen energies $E_n$ are found from the $l$-dimensional Schr\"{o}dinger equation:
\begin{equation}\label{Schrodinger1}
-\frac{\hbar^2}{2m_p}{\bm\nabla}_{\bf r}^2\Psi_n({\bf r})+V({\bf r})\Psi_n({\bf r})=E_n\Psi_n({\bf r}),
\end{equation}
with $m_p$ being a mass of the particle and $V({\bf r})$ being an external potential, and ${\bm\nabla}_{\bf r}=\frac{\partial}{\partial{\bf r}}$.

Among these functionals, a very special role is played by R\'{e}nyi $R_{\rho,\gamma}(\alpha)$ \cite{Renyi1,Renyi2} and Tsallis $T_{\rho,\gamma}(\alpha)$ \cite{Tsallis1} entropies\footnote{From a historical point of view, the latter measures should be more correctly called Havrda-Charv\'{a}t-Dar\'{o}czy-Tsallis entropies \cite{Havrda1,Daroczy1}.} whose expressions in the position (subscript $\rho$) and momentum (subscript $\gamma$) spaces are:
\begin{subequations}
\label{Functionals1}
\begin{eqnarray}\label{RenyiX1}
R_{\rho_n}(\alpha)&=&\frac{1}{1-\alpha}\ln\!\left(\int_{\mathbb{R}^l}\rho_n^\alpha({\bf r})d{\bf r}\right)\\
\label{RenyiP1}
R_{\gamma_n}(\alpha)&=&\frac{1}{1-\alpha}\ln\!\left(\!\int_{\mathbb{R}^l}\gamma_n^\alpha({\bf k})d{\bf k}\right)\\
\label{TsallisX1}
T_{\rho_n}(\alpha)&=&\frac{1}{\alpha-1}\left(1-\int_{\mathbb{R}^l}\rho_n^\alpha({\bf r})d{\bf r}\right)\\
\label{TsallisP1}
T_{\gamma_n}(\alpha)&=&\frac{1}{\alpha-1}\left(1-\int_{\mathbb{R}^l}\gamma_n^\alpha({\bf k})d{\bf k}\right).
\end{eqnarray}
\end{subequations}
They attempt to quantify information with the help of a non-negative parameter, $0<\alpha<\infty$, which can be considered as a factor describing the reaction of the system to its deviation from the equilibrium. Equations~\eqref{Functionals1} show that its small, decreasing to zero values treat the outcomes of the random events more and more in the same manner, regardless of their actual occurrence, and in the extreme case $\alpha\rightarrow0$, they all give the equal contribution to the entropies. As a result, if the limits of integration in any of equations~\eqref{Functionals1} are infinite or semi-infinite, the corresponding entropy (provided it exists) does diverge. In the opposite asymptote of the very large parameter, the events with the highest probabilities are the  only important donors to the integrals in equations~\eqref{Functionals1} with the relative weight of the low-probability  happenings being negligibly small. Special case $\alpha=1$ with the help of the l'H\^{o}pital's rule degenerates for both R\'{e}nyi $R_{\rho,\gamma}(\alpha)$ and Tsallis $T_{\rho,\gamma}(\alpha)$ functionals to the celebrated Shannon entropies \cite{Shannon1}:
\begin{subequations}\label{Shannon1}
\begin{eqnarray}\label{Shannon1_X}
S_{\rho_n}&=&-\int_{\mathbb{R}^l}\rho_n({\bf r})\ln\rho_n({\bf r})d{\bf r}\\
\label{Shannon1_P}
S_{\gamma_n}&=&-\int_{\mathbb{R}^l}\gamma_n({\bf k})\ln\gamma_n({\bf k})d{\bf k}.
\end{eqnarray}
\end{subequations}
Dependence $S$ was introduced in 1948 for the random discrete distribution [see equation~\eqref{RenyiTsallisDiscrete1_S} below] during the mathematical analysis of communication as a measure "of information, choice and uncertainty" \cite{Shannon1}. From a quantum-informational point of view, it is a quantitative descriptor of the lack of our knowledge about the corresponding property of the system: the greater (smaller) the value of this functional is, the less (more) information we have about a nano object. Accordingly, the physical meaning of the R\'{e}nyi entropy and parameter $\alpha$ can be construed as follows: the equilibrium distribution corresponds to $\alpha=1$ and any $\alpha\neq1$ is a deviation from it. Then, the R\'{e}nyi entropy is a measure of the sensitivity of the system to the deviation from the equilibrium. If the parameter $\alpha$ is greater than unity, the corresponding entropy decreases, which means that such a configuration provides more information about the object than its equilibrium counterpart. On the other hand, a distortion in the opposite direction, $\alpha<1$, increases the entropy with the corresponding decrease of the available information and in the extreme case $\alpha\rightarrow0$, it reaches its maximal value, which for the infinite or semi-infinite interval leads to a logarithmic divergence, as mentioned above. Within this limit, the integrand in equations~\eqref{Functionals1} is just a flat unit line, which means that, say, for the position entropy $R_\rho(\alpha)$ we know nothing about particle location in space. The rate of change of the entropy with the R\'{e}nyi parameter just shows the sensitivity of the system to the degree of non-equilibricity.

Another particular case of these entropies are Onicescu energies \cite{Onicescu1}:
\begin{subequations}\label{Onicescu1}
\begin{eqnarray}
\label{Onicescu1_X}
O_{\rho_n}&=&\int_{\mathbb{R}^l}\rho_n^2({\bf r})d{\bf r}\equiv e^{-R_{\rho_n}(2)}\equiv1-T_{\rho_n}(2)\\
\label{Onicescu1_P}
O_{\gamma_n}&=&\int_{\mathbb{R}^l}\gamma_n^2({\bf k})d{\bf k}\equiv e^{-R_{\gamma_n}(2)}\equiv1-T_{\gamma_n}(2),
\end{eqnarray}
\end{subequations}
which measure the deviations of the corresponding distribution from the uniformity.

R\'{e}nyi and Tsallis entropies are related as:
\begin{subequations}\label{RenyiTsallisRelation1}
\begin{eqnarray}
\label{RenyiTsallisRelation1_1}
T&=&\frac{1}{\alpha-1}\left[1-e^{(1-\alpha)R}\right]\\
\label{RenyiTsallisRelation1_2}
R&=&\frac{1}{1-\alpha}\ln(1+(1-\alpha)T),
\end{eqnarray}
\end{subequations}
as it directly follows from equations~\eqref{Functionals1}. Both are decreasing functions of parameter $\alpha$. One important difference between them lies in the fact that the R\'{e}nyi entropies are additive (or extensive) whereas the Thallis functionals are non-additive (or non-extensive). Additivity means that if there are two independent events characterized by their probability functions $f({\bf r})$ and $g({\bf r})$, then the following relation holds:
\begin{equation}\label{AdditivityRenyi1}
R_{fg}(\alpha)=R_f(\alpha)+R_g(\alpha),
\end{equation}
which physically means that the total information obtained by the two independent measurements is exactly the same as the sum of the knowledges received from each separate experiment. The postulate of addititivy of the entropy was a cornerstone in R\'{e}nyi's search for the measure that generalizes the Shannon one \cite{Renyi1}. In turn, the Tsallis entropy is only {\em pseudo}-additive:
\begin{equation}\label{AdditivityTsallis1}
T_{fg}(\alpha)=T_f(\alpha)+T_g(\alpha)+(1-\alpha)T_f(\alpha)T_{g}(\alpha).
\end{equation}
Mathematically, this difference is explained by the presence of the logarithm in the expressions for the R\'{e}nyi entropies. As, for example, equation~\eqref{AdditivityTsallis1} shows, the Shannon entropy, similar to its R\'{e}nyi generalization, is an additive functional too:
\begin{equation}\label{AdditivityShannon1}
S_{fg}(\alpha)=S_f(\alpha)+S_g(\alpha).
\end{equation}
Furhter comparative analysis between R\'{e}nyi and Tsallis entropies can be found, e.g., in \cite{Rajagopal1,Jizba1,Bashkirov1}.

It has to be also noted that the functionals for the continuous distributions from equations~\eqref{Functionals1} were obtained from their discrete counterparts; namely, for the discrete set of all $N$ possible events with their probabilities 
\begin{equation}\label{Requirement1}
0\leq p_n\leq1,\quad n=1,2,\ldots, N,
\end{equation}
so that 
\begin{equation}\label{NormP1}
\sum_{n=1}^Np_n=1,
\end{equation}
and one defines the entropies as
\begin{subequations}\label{RenyiTsallisDiscrete1}
\begin{eqnarray}\label{RenyiTsallisDiscrete1_R}
R(\alpha)&=&\frac{1}{1-\alpha}\ln\!\!\left(\sum_{n=1}^Np_n^\alpha\right)\\
\label{RenyiTsallisDiscrete1_T}
T(\alpha)&=&\frac{1}{\alpha-1}\left(1-\sum_{n=1}^Np_n^\alpha\right)\\
\label{RenyiTsallisDiscrete1_S}
S&=&-\sum_{n=1}^Np_n\ln p_n,
\end{eqnarray}
\end{subequations}
which are, obviously, dimensionless. However, it follows from equations~\eqref{RenyiX1},~\eqref{RenyiP1} and~\eqref{Shannon1}, that the R\'{e}nyi and Shannon entropies for the continuous distributions are measured in units of the logarithm of the length whereas their Tsallis counterparts from equations~\eqref{TsallisX1} and ~\eqref{TsallisP1} represent the sum of the dimensionless unity and the quantity measured in units of the distance raised to power $l(1-\alpha)$ or its inverse. Accordingly, their actual numerical values strongly depend on the units in which the calculations are carried out. Nevertheless, the most importantly, the sum of the position and momentum R\'{e}nyi entropies is a scaling-independent dimensionless quantity. The same is true for the Shannon entropies too. In the case of the Tsallis functionals, a dimensional inconsistency of the two factors in equations~\eqref{TsallisX1} and \eqref{TsallisP1} limits their applications {\em per se} suggesting instead to use the forms $1+(1-\alpha)T(\alpha)$, as done, for instance, in the associated uncertainty relation, equation~\eqref{TsallisInequality1}.

Another important difference between the entropies for the continuous and discrete distributions can be shown in the example of the Shannon functionals \cite{Rudnicki1}; namely, the quantity from equation~\eqref{RenyiTsallisDiscrete1_S} under the requirement from equation~\eqref{Requirement1} is always non-negative, but its counterparts from equations~\eqref{Shannon1} can fall below zero if in some region the corresponding density is greater than unity and its contribution to the integral does overweigh that from the interval where $\rho({\bf r})$ or $\gamma({\bf k})$ lies between zero and one \cite{Olendski3,Olendski1,Olendski4}. Historically, R\'{e}nyi and Tsallis functionals were introduced initially for the discrete distributions in the form from equations~\eqref{RenyiTsallisDiscrete1_R} \cite{Renyi1} and \eqref{RenyiTsallisDiscrete1_T} \cite{Havrda1} first of all for the needs of the theory of information only and their physical applications followed (as it was the case with the original Tsallis postulate \cite{Tsallis1}) with a subsequent generalization to the continuous case. In contrast, long before Shannon, the sum from equation~\eqref{RenyiTsallisDiscrete1_S} was well known exclusively in physics; namely, the very term 'entropy' was introduced into thermodynamics by R. Clausius in 1865 as a function of state of the thermal system that determines its irreversible scattering of energy. Classical statistical mechanics based on Boltzmann-Gibbs (BG) approach treats the quantity from equation~\eqref{RenyiTsallisDiscrete1_S} (with its right-hand side multiplied by the Boltzmann constant $k_B$) as a measure of disorder with $p_n$ describing  the probability of the microstate with the energy $E_n$ to occur and with the total number of the microscopic configurations in the system being $N$ whereas the quantum consideration replaces it by von Neumann entropy,
\begin{equation}\label{vonNeumann1}
S=-k_B{\rm Sp}(\hat{\rho}\ln\hat{\rho})
\end{equation}
with $\hat{\rho}$ being the density matrix operator. A pivotal role of the BG entropy can be shown on the example of the canonical distribution when a condition of maximization of $S$ with the additional requirement of the total internal energy 
\begin{equation}\label{InternalEnergy1}
U=\sum_{n=1}^Np_nE_n
\end{equation}
staying unchanged leads to the Gibbs distribution. Tsallis \cite{Tsallis1} was the first to point out that the entropy from equation~\eqref{RenyiTsallisDiscrete1_T} does generalize the BG statistics; namely, its extremization with the help of the Lagrange parameters yields:
\begin{subequations}\label{Extreme1}
\begin{eqnarray}\label{Extreme1_p}
p_n(\alpha;T_t)&=&\frac{[1-(\alpha-1)E_n/(k_BT_t)]^{1/(\alpha-1)}}{Z(\alpha;T_t)},\\
\label{Extreme1_Z}
Z(\alpha;T_t)&=&\sum_{n=1}^N[1-(\alpha-1)E_n/(k_BT_t)]^{1/(\alpha-1)},
\end{eqnarray}
\end{subequations}
where $T_t$ is an absolute thermodynamic temperature with subindex 't' distinguishing it from the Tsallis entropy $T(\alpha)$. It can be immediately seen that the limit $\alpha\rightarrow1$ recovers the BG expressions:
\begin{subequations}\label{Extreme2}
\begin{eqnarray}
p_n(\alpha;T_t)=\frac{1}{Z(1;T_t)}e^{-E_n/(k_BT_t)}\nonumber\\
\label{Extreme2_p}
\times\!\!\left(\!1\!-\!\frac{1}{2}\!\!\left[\!\left(\!\frac{E_n}{k_BT_t}\!\right)^{\!\!\!2}-\frac{1}{Z(1;T_t)}\sum_{n'=1}^N\!\left(\!\frac{E_{n'}}{k_BT_t}\!\right)^{\!\!\!2}\!e^{-E_{n'}/(k_BT_t)}\right]\!\!(\alpha-1)+\ldots\right),\\
\label{Extreme2_Z}
Z(\alpha;T_t)=\sum_{n=1}^Ne^{-E_n/(k_BT_t)}-\frac{1}{2}(\alpha-1)\sum_{n=1}^N\left(\!\frac{E_n}{k_BT_t}\!\right)^{\!\!\!2}e^{-E_n/(k_BT_t)}+\ldots,
\end{eqnarray}
\end{subequations}
where
\begin{equation}\label{Partition1}
Z(1;T_t)\equiv Z(T_t)=\sum_{n=1}^Ne^{-E_n/(k_BT_t)}
\end{equation}
is a BG partition function. In this way, Tsallis statistics supplements the BG one by expanding it to $\alpha\neq1$. Non-Gaussian distributions from equation~\eqref{Extreme1_p} have been predicted theoretically \cite{Lutz1} and successfully applied for the explanation of the experiments in miscellaneous branches of physics \cite{Douglas1,Liu1,Pickup1}, including high-energy collisions \cite{Tang1,Shao1}. For understanding thermodynamic interpretation of the R\'{e}nyi entropy \cite{Beck1,Baez1,Luitz1,Mora1}, one needs to substitute the BG distribution $p_n(1;T_1)$ from equation~\eqref{Extreme2_p} at the temperature $T_1$ into equation~\eqref{RenyiTsallisDiscrete1_R} (right-hand side of which, similar to the Shannon entropy, has to be multiplied by $k_B$) and represent the R\'{e}nyi parameter as the ratio of the two temperatures, $\alpha=T_1/T_2$, arriving in this way at
\begin{equation}\label{RenyiFreeEnergy1}
R\!\left(\frac{T_1}{T_2}\right)=-\frac{F(T_2)-F(T_1)}{T_2-T_1},
\end{equation}
where $F(T_t)$ is Helmholtz free energy \cite{Helrich1}:
\begin{equation}\label{RenyiFreeEnergy2}
F(T_t)=-k_BT_t\ln Z(T_t).
\end{equation}
Thus, the R\'{e}nyi entropy replaces the derivative of the free energy in the well-known thermodynamic relation between $S$ and $F$ \cite{Helrich1}:
\begin{equation}\label{EntropyFreeEnergy1}
S(T_t)=-\frac{\partial F}{\partial T_t},
\end{equation}
by the ratio of the finite differences at the two temperatures. Drawing a parallel with the velocities in mechanics \cite{Walker1}, one can say that the Shannon entropy describes the instantaneous speed of change of the Helmholtz energy with temperature whereas the R\'{e}nyi functional averages this variation (and, accordingly, the amount of work the system can do) over some interval $T_2-T_1$.

For any quantum orbital, the position and momentum components of the entropies are not independent from each other; for example, the R\'{e}nyi entropies obey the following fundamental inequality \cite{Bialynicki1,Zozor1}:
\begin{equation}\label{RenyiUncertainty1}
R_{\rho_n}(\alpha)+R_{\gamma_n}(\beta)\geq-\frac{l}{2}\left(\frac{1}{1-\alpha}\ln\frac{\alpha}{\pi}+\frac{1}{1-\beta}\ln\frac{\beta}{\pi}\right)
\end{equation}
with the constraint on the interrelation between the indexes:
\begin{equation}\label{RenyiUncertainty2}
\frac{1}{\alpha}+\frac{1}{\beta}=2.
\end{equation}
Within the limit of $\alpha\rightarrow1$,  equation~\eqref{RenyiUncertainty1} degenerates to the Shannon uncertainty relation \cite{Bialynicki2,Beckner1}:
\begin{equation}\label{ShannonInequality}
S_{\rho_n}+S_{\gamma_n}\geq l(1+\ln\pi).
\end{equation}
Inequality similar to equation~\eqref{RenyiUncertainty1} exists for the Tsallis entropies too \cite{Rajagopal1}:
\begin{equation}\label{TsallisInequality1}
\left(\frac{\alpha}{\pi}\right)^{l/(4\alpha)}\!\!\left[1+(1-\alpha)T_{\rho_n}(\alpha)\right]^{1/(2\alpha)}\geq\left(\frac{\beta}{\pi}\right)^{l/(4\beta)}\!\!\left[1+(1-\beta)T_{\gamma_n}(\beta)\right]^{1/(2\beta)}, \end{equation}
what is a direct consequence of the Sobolev inequality of the Fourier transform \cite{Beckner2}:
\begin{equation}\label{Sobolev1}
\left(\frac{\alpha}{\pi}\right)^{l/(4\alpha)}\left[\int_{\mathbb{R}^l}\rho_n^\alpha({\bf r})d{\bf r}\right]^{1/(2\alpha)}\geq\left(\frac{\beta}{\pi}\right)^{l/(4\beta)}\left[\int_{\mathbb{R}^l}\gamma_n^\beta({\bf k})d{\bf k}\right]^{1/(2\beta)},
\end{equation}
in the proof of which, in addition to the requirement from equation~\eqref{RenyiUncertainty2}, an extra restriction
\begin{equation}\label{Sobolev2}
\frac{1}{2}\leq\alpha\leq1
\end{equation}
is imposed \cite{Beckner2}. As immediately follows from equations~\eqref{TsallisInequality1} and \eqref{Sobolev1}, they are saturated at $\alpha=\beta=1$ when either side turns to a dimensionless $\pi^{-l/4}$. Observe also that R\'{e}nyi inequality~\eqref{RenyiUncertainty1} is obtained by taking the logarithms of its Sobolev counterpart, equation~\eqref{Sobolev1}, and it is extremely important to point out that as a result of this operation, the condition from equation~\eqref{Sobolev2} is removed for the R\'{e}nyi entropies \cite{Zozor1}. This difference between the two uncertainty relations, equations~\eqref{RenyiUncertainty1} and \eqref{TsallisInequality1}, will be illustrated below in the particular examples.

R\'{e}nyi and Tsallis entropies find countless applications in science, technology, engineering, medicine, economics, among others; for example, soon after a Hungarian mathematician proposed the functional now bearing his name \cite{Renyi1}, R\'{e}nyi measure was successfully applied in coding theory \cite{Campbell1}. Its applications in physics include but not are limited to the analysis of the processes of hadronic multiparticle high-energy collisions \cite{Bialas1,Bialas2}; fractional diffusion processes \cite{Essex1}; properties of the XY spin chain \cite{Franchini1}; the flow to the system in thermal equilibrium that is weakly coupled to an arbitrary system out of equilibrium subject to arbitrary time-depending forces \cite{Ansari1}; conformal field theory \cite{Klebanov1,Chen1}; black hole area law \cite{Dong1}; and many, many others (see, for example, sources in \cite{Klebanov1,Chen1,Dong1}). Recent state-of-the-art experiments directly measured R\'{e}nyi entropy \cite{Islam1}, which opens up new horizons in using quantum entanglement to characterize the dynamics of strongly correlated many-body systems. For our subsequent discussion, one has to note that R\'{e}nyi, equation~\eqref{RenyiUncertainty1}, Shannon, equation~\eqref{ShannonInequality}, and Tsallis, equation~\eqref{TsallisInequality1}, inequalities are more general in establishing a relationship between the two non-commuting operators than the standard Heisenberg uncertainty relation that for the one-dimensional (1D) geometry, $l=1$, we will write as:
\begin{equation}\label{Heisenberg1}
\Delta x\Delta k\geq\frac{1}{2}.
\end{equation}
Here, $\Delta x$ and $\Delta k$ are, respectively, position and wave vector standard deviations:
\begin{subequations}\label{DeltaXK1_1}
\begin{eqnarray}\label{DeltaX1_1}
\Delta x&=&\sqrt{\langle x^2\rangle-\langle x\rangle^2},\\
\label{DeltaK1_1}
\Delta k&=&\sqrt{\langle k^2\rangle-\langle k\rangle^2},
\end{eqnarray}
\end{subequations}
where the associated moments $\langle x^j\rangle$ and $\left<k^j\right>$, $j=1,2,\ldots$, are expressed with the help of the corresponding position $\rho(x)$ and momentum $\gamma(k)$ densities:
\begin{subequations}\label{XKaveraging1}
\begin{eqnarray}\label{Xaveraging1}
\langle x^j\rangle&=&\int_{\mathbb{R}^1}x^j\rho(x)dx\\
\label{Kaveraging1}
\langle k^j\rangle&=&\int_{\mathbb{R}^1}k^j\gamma(k)dk.
\end{eqnarray}
\end{subequations}
To show the non-universality of the Heisenberg relation, one might consider the lowest level of the 1D Neumann well of width $a$: its energy is zero and its position waveform is just a constant $\Psi_0^N(x)=a^{-1/2}$, $-a/2\leq x\leq a/2$. Then, its momentum wave function reads:
\begin{equation}\label{Neumann1DMomentum1}
\Phi_0^N(k)=\left(\frac{2}{\pi a}\right)^{1/2}\frac{1}{k}\sin\frac{ka}{2},\quad-\infty\leq k\leq+\infty.
\end{equation}
From this expression, it can immediately be seen that the the second moment $\langle k^2\rangle$ and, together with it, dispersion diverge: $\Delta k=\infty$ \cite{Uffink1,Majernik1}, which makes the Heisenberg relation meaningless since it does not bring about any new information about position-momentum limitations: infinity is always greater than any finite number. If, instead of defining momentum moments as it is done in equation~\eqref{Kaveraging1}, one uses alternately the wave vector operator $\hat{k}=-i\partial/\partial x$ acting upon the space of the position functions:
\begin{equation}\tag{31b$'$}\label{eq:31b'}
\langle k^j\rangle=\int_{\mathbb{R}^1}\Psi(x)\hat{k}^j\Psi(x)dx,
\end{equation}
then it can be seen that for the Neumann ground level the Heisenberg relation is violated since the momentum dispersion turns to zero, $\Delta k=0$ \cite{Olendski3}. However, both R\'{e}nyi \cite{Bialynicki3,Bialynicki4}, equation~\eqref{RenyiUncertainty1}, and Shannon \cite{Bialynicki3,Bialynicki4,Majernik1,Olendski3}, equation~\eqref{ShannonInequality}, inequalities hold true with, for example, the left-hand side of the latter expression yielding $2.6834\ldots$ \cite{Olendski3} whereas  $1+\ln\pi=2.1447\ldots$.

Returning to the applications of the entropies, let us mention that, besides physics and information theory, R\'{e}nyi measure was used in the investigation of spatial distribution of earthquake epicentres \cite{Geilikman1}, for the analysis of the landscape diversity and integrity \cite{Carranza1,DeLuca1,Drius1,Rocchini1}, the examination of the behaviour of the stock markets \cite{Jizba2,Jizba3}; characterization of scalp electroencephalogram records corresponding to secondary generalized tonic-clonic epileptic seizures \cite{Rosso1}; study of the biological signals \cite{Costa1}, exploration and modification of the brain activity \cite{Tozzi1}; in digital image analysis \cite{Peters1}, etc. In turn, Tsallis entropies are widely employed in non-extensive systems, which include the structures and processes characterized by non-ergodicity, long-range correlations and space-time (multi)fractal geometry; see, e.g., reference \cite{Tsallis2} for more details. The literature of both entropies is growing at an impressive rate.

Despite a paramount significance of these two entropies, concrete results on the specific physical quantum systems to date are scarce, which is probably explained by the tremendous difficulties of evaluating (even numerically) the integrals entering them, especially for the momentum components. By means of semi-classical approximation, the general asymptotic formulae were derived for the 1D position R\'{e}nyi entropy and checked for the infinite potential well \cite{SanchezRuiz1}. For the same structure the expressions for the entropies $R_{\gamma_n}(\alpha)$ and $T_{\gamma_n}(\alpha)$, which contain double finite sums, were provided for the integer values of the parameter $\alpha$ only \cite{Aptekarev1}. The entropies of the highly excited levels for the 3D hydrogenic system \cite{Toranzo1} and harmonic oscillator (HO) \cite{Dehesa1} were scrutinized too and it was shown, in particular, that the ground orbital of the latter structure saturates the entropic relation, equation~\eqref{RenyiUncertainty1}. The results were generalized to any number of dimensions \cite{Aptekarev2,PuertasCenteno1}.

In view of these remarks, it is extremely important to investigate the quantum structures where {\em analytic} expressions for $R$ and $T$ are possible.  Below, we consider three such 1D systems. They are:
\begin{itemize}
\item{HO\\}
\item{attractive Robin wall\\}
and
\item{quasi-1D (Q1D) hydrogenic atom.}
\end{itemize}
For the second structure, only one bound state exists whereas the first and the last geometries are characterized by the infinite number of localized levels. For the HO, analytic expressions of the entropies can be derived for the ground and first excited orbitals and for the Q1D hydrogenic atom exact results in terms of the $\Gamma$-function \cite{Abramowitz1} are derived for the momentum components of any orbital and for the position part of the ground state. Obtained formulae for the entropies allow a simple analysis of the asymptotic cases; in particular, it is explicitly shown that the position components do exist at any non-negative R\'{e}nyi parameter whereas their momentum counterparts are valid for the range of the index $\alpha$, which is limited from below by the positive threshold $\alpha_{TH}>0$. This lower bound on the {\em momentum} entropies strongly depends on the {\em position} potential and the type of the boundary condition imposed on the \emph{position} waveform; for example, it is equal to one half for the Robin wall, $\alpha_{TH}^{RW}=1/2$, and to one quarter for the hydrogenic atom, $\alpha_{TH}^{Q1D}=1/4$. It is shown that the entropic uncertainty relation, equation~\eqref{RenyiUncertainty1}, is always satisfied and is saturated at the limiting values of the parameters from equation~\eqref{RenyiUncertainty2}. Since the results below are easily comprehensible (both mathematically as well as physically) by the students, they can serve as part of a graduate introduction to quantum-information theory. The previous pedagogical approach to studying quantum-information measures \cite{Saha1} addressed Fisher informations \cite{Fisher1,Frieden1} of the HO, infinitely deep Dirichlet well and Q1D hydrogen atom. Unfortunately, that research suffered a severe blunder in the analysis of the momentum waveform of the latter structure \cite{Olendski2}. We shall revisit this issue in more detail in chapter~\ref{Sec_Q1DHA}. As a result of our analysis, we point out at some appropriate points that require further consideration.
\section{1D HO}\label{sec_HO}
\begin{figure}
\centering
\includegraphics[width=\columnwidth]{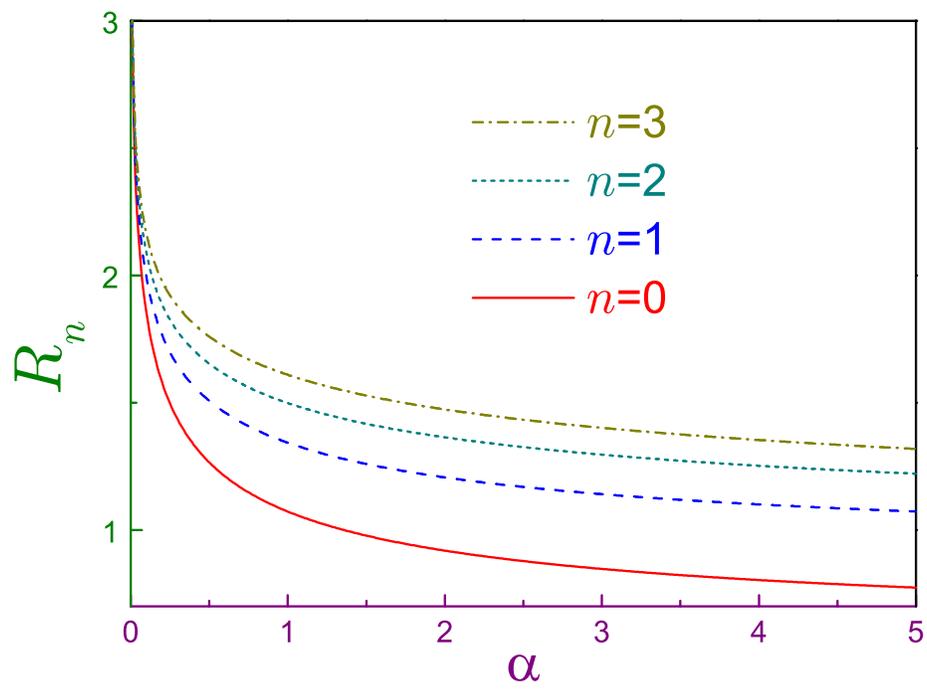}
\caption{\label{HOFig1}
R\'{e}nyi entropies $R_n(\alpha)$ of the four lowest HO states as functions of parameter $\alpha$.}
\end{figure}

Consider a motion along the whole $x$ axis of the quantum particle in the potential of the form:
\begin{equation}\label{PotentialHO1}
V(x)=\frac{1}{2}\,m_p\omega^2x^2.
\end{equation}
Then, eigen energies $E_n$ and associated position eigen functions $\Psi_n(x)$ of the 1D Schr\"{o}dinger equation take the form \cite{Landau1}:
\begin{eqnarray}\label{HO_Energies1}
E_n=\hbar\,\omega\!\left(n+\frac{1}{2}\right),\\
\label{HO_Psi1}
\Psi_n(x)=\frac{1}{r_\omega^{1/2}}\frac{1}{\pi^{1/4}}\frac{1}{\left(2^nn!\right)^{1/2}}\,e^{-(x/r_\omega)^2/2}H_n\!\left(\frac{x}{r_\omega}\right),
\end{eqnarray}
$n=0,1,\ldots$. Here, $H_n(x)$ is $n$th order Hermite polynomial \cite{Abramowitz1,Lebedev1}, and $r_\omega=\left[\hbar/\left(m_p\omega\right)\right]^{1/2}$. Corresponding momentum waveforms can also be easily obtained from equation \eqref{Fourier1_1} employing, for example, properties of the generating function of the Hermite polynomials \cite{Lebedev1}:
\begin{equation}\label{HO_Phi1}
\Phi_n(k)=(-i)^nr_\omega^{1/2}\frac{1}{\pi^{1/4}}\frac{1}{\left(2^nn!\right)^{1/2}}\,e^{-(r_\omega k)^2/2}H_n\left(r_\omega k\right).
\end{equation}
An alternative equivalent way of arriving at the same expression is to use a momentum representation of the Schr\"{o}dinger equation when the coordinate $x$ is an operator, $\hat{x}=id/dk$ \cite{Landau1}. A comparison of equations \eqref{HO_Psi1} and \eqref{HO_Phi1} reveals that the position $\rho_n(x)$ and momentum $\gamma_n(k)$ densities are related as
\begin{equation}\label{HO_relation1}
\gamma_n(k)=r_\omega^2\rho_n\left(r_\omega^2k\right).
\end{equation}
As a result, corresponding R\'{e}nyi entropies read:
\begin{subequations}\label{HO_Renyi1}
\begin{eqnarray}\label{HO_RenyiX1}
R_{\rho_n}(\alpha)=\ln r_\omega+\frac{1}{1-\alpha}\ln I_n(\alpha)\\
\label{HO_RenyiK1}
R_{\gamma_n}(\alpha)=-\ln r_\omega+\frac{1}{1-\alpha}\ln I_n(\alpha)
\end{eqnarray}
\end{subequations}
with
\begin{equation}\label{HO_Integral1}
I_n(\alpha)=\frac{1}{\left(\pi^{1/2}2^nn!\right)^\alpha}\int_{-\infty}^\infty\left[e^{-z^2}H_n^2(z)\right]^\alpha\!dz.
\end{equation}
It can immediately be seen that the sum of the two entropies from the left-hand side of inequality \eqref{RenyiUncertainty1} is a dimensionless scale-independent quantity, as it should be. Accordingly, below in this section, while discussing the quantities $R_\rho$ and $R_\gamma$, we will use the units in which the length is measured in terms of $r_\omega$, which means that the HO position and momentum R\'{e}nyi entropies are equal to each other:
\begin{equation}\label{HO_relation2}
R_{\rho_n}(\alpha)\equiv R_{\gamma_n}(\alpha).
\end{equation}
Then, if it will not cause any confusion, we will drop the subindex $\rho$ or $\gamma$. Remarkably, for the ground, $n=0$, and first excited, $n=1$, states these quantities are evaluated analytically:
\begin{subequations}\label{HO_R01}
\begin{eqnarray}\label{HO_R01_0}
R_0(\alpha)=\frac{1}{2}\ln\pi-\frac{1}{2}\frac{\ln\alpha}{1-\alpha}\\
\label{HO_R01_1}
R_1(\alpha)=\frac{1}{1-\alpha}\ln\!\left(\frac{2^\alpha}{\pi^{\alpha/2}\alpha^{\alpha+1/2}}\Gamma\!\left(\alpha+\frac{1}{2}\right)\right).
\end{eqnarray}
\end{subequations}
They are plotted in figure~\ref{HOFig1} together with their $n=2$ and $n=3$ counterparts, which can be evaluated only numerically. It can be seen that, as expected, the R\'{e}nyi entropies are decreasing functions of parameter $\alpha$ and for the higher-lying states they are greater than their lower-lying counterparts. At the vanishing $\alpha$ all the entropies diverge logarithmically:
\begin{subequations}\label{HO_R01_Asymptote0}
\begin{eqnarray}\label{HO_R01_Asymptote0_0}
R_0(\alpha)=\frac{1}{2}\left[\ln\pi-(1+\alpha)\ln\alpha\right],\quad\alpha\rightarrow0,\\
\label{HO_R01_Asymptote0_1}
R_1(\alpha)=\frac{1}{2}(\ln\pi-\ln\alpha)-\left[\gamma+\ln\left(2\pi^{1/2}\right)+\ln\alpha\right]\alpha,\quad\alpha\rightarrow0,
\end{eqnarray}
\end{subequations}
which is explained  by the infinite limits of integration of the constant unit function, as mentioned in the Introduction, and at the huge R\'{e}nyi parameter, $\alpha\rightarrow\infty$, for the two lowest states they are:
\begin{subequations}\label{HO_R01_AsymptInfinite}
\begin{eqnarray}\label{HO_R01_AsymptInfinite_0}
R_0(\alpha)=\frac{1}{2}\ln\pi+\frac{1}{2}\frac{\ln\alpha}{\alpha}+\ldots,\\
\label{HO_R01_AsymptInfinite_1}
R_1(\alpha)=1-\ln2+\frac{1}{2}\ln\pi+\frac{1}{2}\frac{\ln\alpha-3\ln2+2}{\alpha}+\ldots,
\end{eqnarray}
\end{subequations}
with the leading terms in these asymptotic expansions being $\frac{1}{2}\ln\pi=0.5723\ldots$ and $1-\ln2+\frac{1}{2}\ln\pi=0.8792\ldots$, respectively, while in the vicinity of the Shannon case, $\alpha\rightarrow1$, they turn to
\begin{subequations}\label{HO_R01_Asympt1}
\begin{eqnarray}\label{HO_R01_Asympt1_0}
R_0(\alpha)&=\frac{1}{2}(1+\ln\pi)-\frac{1}{4}(\alpha-1)+\frac{1}{6}(\alpha-1)^2+\ldots\\
R_1(\alpha)&=\ln2+\gamma+\frac{1}{2}\ln\pi-\frac{1}{2}-\frac{\pi^2-9}{4}(\alpha-1)\nonumber\\
\label{HO_R01_Asympt1_1}
&+\frac{7\zeta(3)-8}{3}(\alpha-1)^2+\ldots.
\end{eqnarray}
\end{subequations}
Here, $\gamma$ is Euler's constant \cite{Abramowitz1}:
\begin{equation}\label{Euler1}
\gamma=\lim_{n\rightarrow\infty}\left(\sum_{i=1}^n\frac{1}{i}-\ln n\right)=0.5772\ldots
\end{equation}
and $\zeta(z)$ is Riemann zeta function \cite{Abramowitz1}:
\begin{equation}\label{Zeta1}
\zeta(z)=\sum_{k=1}^\infty\frac{1}{k^z}
\end{equation}
with $\zeta(3)=1.2020\ldots$. Accordingly, the leading terms in equations~\eqref{HO_R01_Asympt1} are $1.0723\ldots$ and $1.3427\ldots$. For our subsequent analysis, it is important to point out that the momentum entropies are defined for any positive R\'{e}nyi parameter $\alpha$.

\begin{figure}
\centering
\includegraphics[width=\columnwidth]{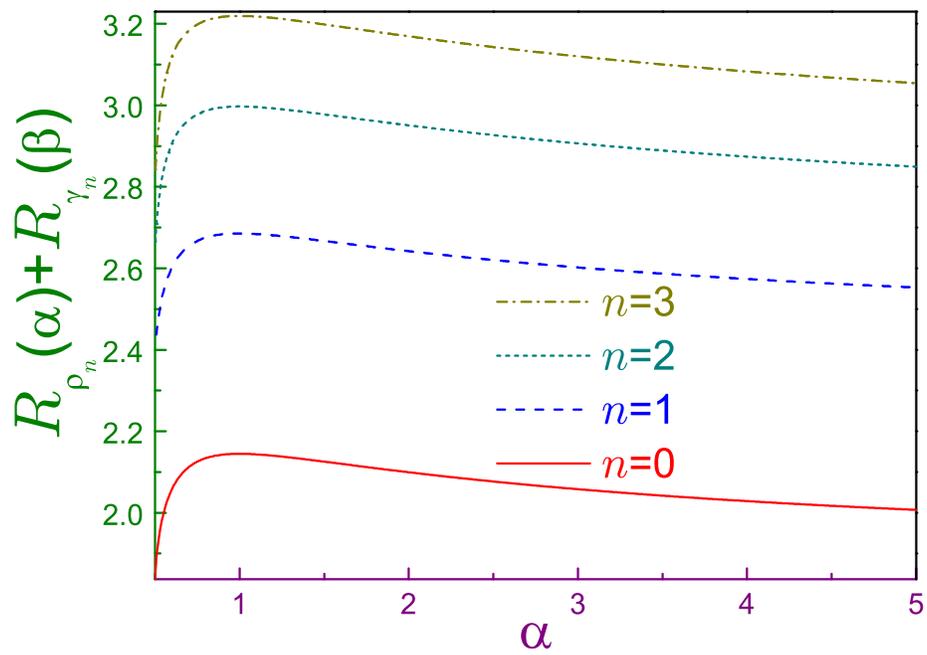}
\caption{\label{HOFig2}
Sum of the position and momentum R\'{e}nyi entropies $R_{\rho_n}(\alpha)+R_{\gamma_n}(\beta)$ with parameter $\beta$ from equation~\eqref{beta1} of the four lowest HO states as functions of parameter $\alpha$.}
\end{figure}

To get deeper insight into the structure of these entropies, it is worthwhile, by using relation~\eqref{RenyiUncertainty2}, to bring inequality \eqref{RenyiUncertainty1} to the following form \cite{Zozor1}:
\begin{equation}\label{RenyiUncertainty3}
R_{\rho_n}(\alpha)+R_{\gamma_n}(\beta)\geq\ln\pi-\frac{1}{2}\left(\frac{\ln\alpha}{1-\alpha}+\frac{\ln\beta}{1-\beta}\right).
\end{equation}
Then, one can see that the Gaussian probability distribution, i.e., ground-level, $n=0$, densities with $\rho_0(x)=\pi^{-1/2}e^{-x^2}$ and  $\gamma_0(k)=\pi^{-1/2}e^{-k^2}$, do saturate the uncertainty relation at any $\alpha$ and $\beta$ that obey restriction from equation~\eqref{RenyiUncertainty2}, as it was pointed out before \cite{Bialynicki1}. Denoting right-hand-side of inequalities~\eqref{RenyiUncertainty1} and \eqref{RenyiUncertainty3} where, as it follows from equation \eqref{RenyiUncertainty2}, parameter $\beta$ is a function of $\alpha$:
\begin{equation}\label{beta1}
\beta=\frac{\alpha}{2\alpha-1},
\end{equation}
by $f(\alpha)$:
\begin{equation}\label{Function_f1}
f(\alpha)=\ln\pi-\left[\ln\alpha-\frac{\alpha-1/2}{\alpha-1}\ln(2\alpha-1)\right],
\end{equation}
it is easy to get its asymptotic limits:
\begin{subequations}\label{HO_Flimits}
\begin{eqnarray}\label{HO_FlimitsOneHalf}
f(\alpha)=\ln2\pi-[1+\ln(2\alpha-1)](2\alpha-1)+\ldots,\quad\alpha\rightarrow\frac{1}{2}\\
\label{HO_FlimitsOne}
f(\alpha)=1+\ln\pi-\frac{1}{6}\,(\alpha-1)^2+\frac{1}{3}\,(\alpha-1)^3+\ldots,\quad\alpha\rightarrow1\\
\label{HO_FlimitsInfinity}
f(\alpha)=\ln2\pi+\frac{\ln2\alpha-1}{2\alpha}+\ldots,\quad\alpha\rightarrow\infty
\end{eqnarray}
\end{subequations}
with $\ln2\pi=1.8378\ldots$ and numerical value of $1+\ln\pi$ provided in the Introduction. Equation~\eqref{HO_FlimitsOne} manifests that the Shannon case, $\alpha=1$, presents a maximum of the function $f(\alpha)$ and, accordingly, of $R_{\rho_0}(\alpha)+R_{\gamma_0}(\beta)$. As the solid curve in figure~\ref{HOFig2} depicts, it is its global extremum. Observe that equations~\eqref{HO_FlimitsOneHalf} and \eqref{HO_FlimitsInfinity} turn into each other under the transformation
\begin{equation}\label{AlphaTransform1}
\alpha\rightarrow\beta=\alpha/(2\alpha-1),
\end{equation}
which is a reflection of the symmetry of the system upon introduction of the symmetrized parameter $s,$ so that $\alpha=1/(1-s)$, $\beta=1/(1+s)$, $-1\leq s\leq+1$, as it was done in reference~\cite{Bialynicki1}. With the help of equation~\eqref{HO_R01_1} one can also find the corresponding limits of the first excited state:
\begin{subequations}\label{HO_limitsR1}
\begin{eqnarray}
R_{\rho_1}(\alpha)+R_{\gamma_1}(\beta)=1+\ln4\nonumber\\\
\label{HO_limitsR1OneHalf}
-2\left[\gamma+\ln\pi+\ln\left(\alpha-\frac{1}{2}\right)\right]\!\!\left(\alpha-\frac{1}{2}\right)+\ldots,\quad\alpha\rightarrow\frac{1}{2},\\
R_{\rho_1}(\alpha)+R_{\gamma_1}(\beta)=2\gamma-1+\ln4\pi\nonumber\\
\label{HO_limitsR1One}
+\left[\frac{14}{3}\zeta(3)-\frac{\pi^2}{2}-\frac{5}{6}\right]\,(\alpha-1)^2+\ldots,\quad\alpha\rightarrow1,\\
\label{HO_limitsR1Infinity}
R_{\rho_1}(\alpha)+R_{\gamma_1}(\beta)=1+\ln4+\frac{\ln\frac{4}{\pi}-\gamma+\ln\alpha}{2\alpha}+\ldots,\quad\alpha\rightarrow\infty.
\end{eqnarray}
\end{subequations}
First, note that since $\frac{14}{3}\zeta(3)-\frac{\pi^2}{2}-\frac{5}{6}=-0.1585\ldots$ is negative, the Shannon case of $\alpha=1$ is again a maximum of the sum of the two R\'{e}nyi entropies. Next, equations~\eqref{HO_limitsR1OneHalf} and \eqref{HO_limitsR1Infinity}, similar to equations~\eqref{HO_FlimitsOneHalf} and \eqref{HO_FlimitsInfinity}, transform into each other under the conversion from equation~\eqref{AlphaTransform1}. Third, the leading terms in equations~\eqref{HO_limitsR1}, namely, $1+\ln4=2.3862\ldots$ and $2\gamma-1+\ln4\pi=2.6854\ldots$, are greater than their ground-orbital counterparts from equations~\eqref{HO_Flimits}. This remains true for any positive $\alpha$. These properties, i.e., the increase of the sum $R_{\rho_n}(\alpha)+R_{\gamma_n}(\beta)$ with the quantum index $n$ and its global extremum at the Shannon entropy holds for any excited level, as figure~\ref{HOFig2} demonstrates where the values for $n=2$ and $n=3$ were computed numerically. Let us also point out that the R\'{e}nyi inequalities hold true for any $\alpha$ greater than one half, without additional restriction from equation~\eqref{Sobolev2}.

\begin{figure}
\centering
\includegraphics[width=\columnwidth]{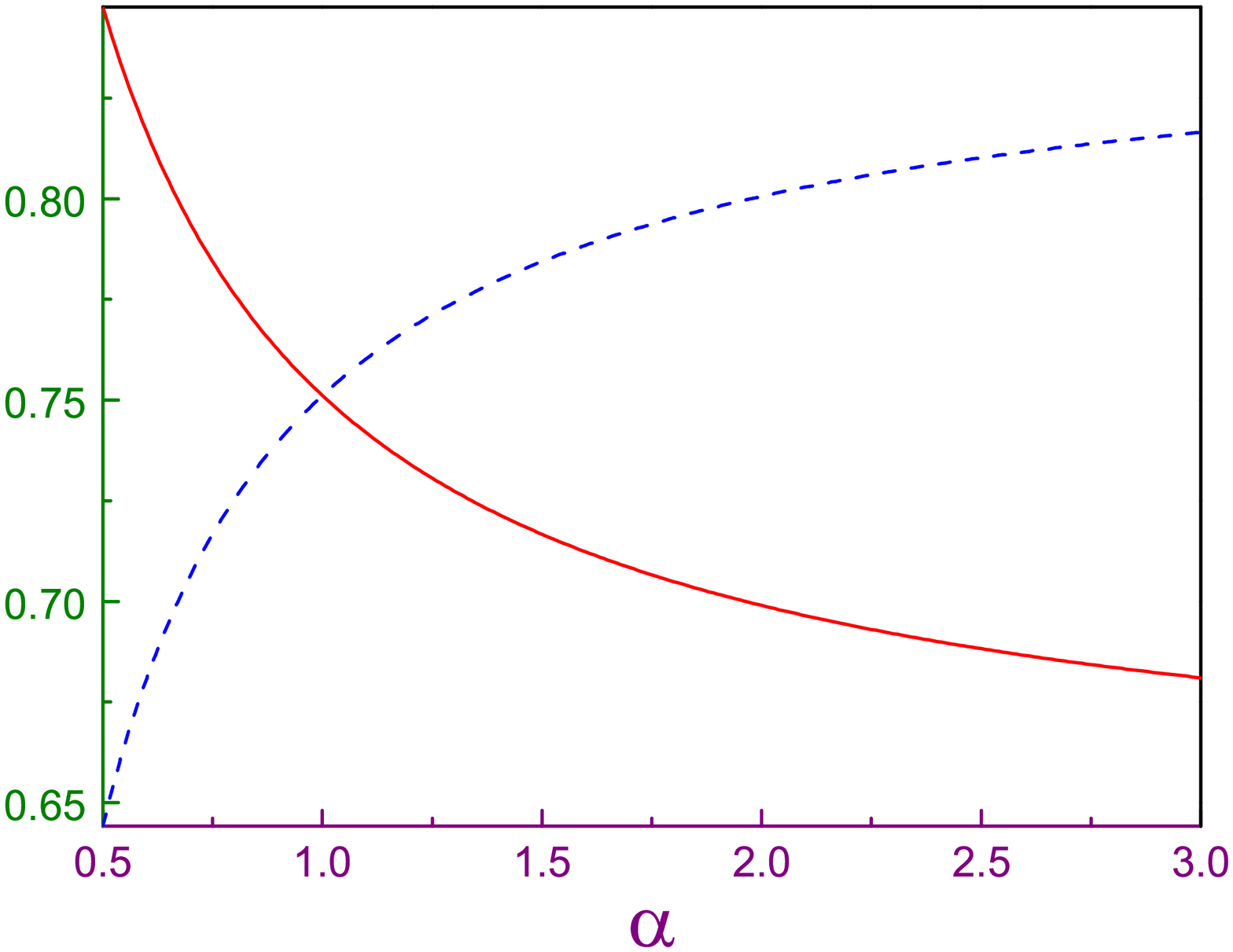}
\caption{\label{HOTsallisFig1}
Left- (solid line) and right- (dashed curve) hand sides of relation~\eqref{HO_Tsallis2} as functions of the parameter $\alpha$ where it is assumed that $r_\omega\equiv1$.}
\end{figure}

Tsallis entropies for the two lowest lying states read:
\begin{subequations}\label{HO_Tsallis1}
\begin{eqnarray}\label{HO_Tsallis1_X0}
T_{\rho_0}(\alpha)=\frac{1}{\alpha-1}\left[1-\frac{1}{\alpha^{1/2}\left(\pi^{1/2}r_\omega\right)^{\alpha-1}}\right]\\
\label{HO_Tsallis1_K0}
T_{\gamma_0}(\alpha)=\frac{1}{\alpha-1}\left[1-\frac{1}{\alpha^{1/2}}\left(\frac{r_\omega}{\pi^{1/2}}\right)^{\alpha-1}\right]\\
\label{HO_Tsallis1_X1}
T_{\rho_1}(\alpha)=\frac{1}{\alpha-1}\left[1-\frac{2^\alpha}{\pi^{\alpha/2}r_\omega^{\alpha-1}}\frac{\Gamma\!\left(\alpha+\frac{1}{2}\right)}{\alpha^{\alpha+1/2}}\right]\\
\label{HO_Tsallis1_K1}
T_{\gamma_1}(\alpha)=\frac{1}{\alpha-1}\left[1-\frac{2^\alpha}{\pi^{\alpha/2}}\frac{\Gamma\!\left(\alpha+\frac{1}{2}\right)}{\alpha^{\alpha+1/2}}r_\omega^{\alpha-1}\right]
\end{eqnarray}
\end{subequations}
where we switched back to the regular dimensional units. First, let us point out that taking the limit $\alpha\rightarrow1$ of, for example, equations~\eqref{HO_Tsallis1_X0} and \eqref{HO_Tsallis1_K0}:
\begin{subequations}\label{HO_Tsallis1a}
\begin{eqnarray}
T_{\rho_0}(\alpha)=&\ln r_\omega+\frac{1}{2}(1+\ln\pi)\nonumber\\
\label{HO_Tsallis1a_X0}
&-\frac{1}{2}\left[\ln^2\!\left(\pi^{1/2}r_\omega\right)+\ln\left(\pi^{1/2}r_\omega\right)+\frac{3}{4}\right](\alpha-1)+\ldots\\
T_{\gamma_0}(\alpha)=&-\ln r_\omega+\frac{1}{2}(1+\ln\pi)\nonumber\\
\label{HO_Tsallis1a_K0}
&-\frac{1}{2}\left[\ln^2\!\left(\frac{\pi^{1/2}}{r_\omega}\right)+\ln\left(\frac{\pi^{1/2}}{r_\omega}\right)+\frac{3}{4}\right](\alpha-1)+\ldots,
\end{eqnarray}
\end{subequations}
and comparing them with equation~\eqref{HO_R01_Asympt1_0} (where, as we remember, the convention $r_\omega\equiv1$ was used), one sees that R\'{e}nyi and Tsallis entropies approach their Shannon asymptote in different ways. The same can be shown for the first exiting level, equations~\eqref{HO_Tsallis1_X1} and \eqref{HO_Tsallis1_K1}, but since the resulting expressions are bulky, they are not provided here. Next, plugging in equations~\eqref{HO_Tsallis1_X0} and \eqref{HO_Tsallis1_K0} into the entropy inequality~\eqref{TsallisInequality1}, one calculates its left- and right-hand sides as $\pi^{-1/4}r_\omega^{\frac{1-\alpha}{2\alpha}}$ and $\pi^{-1/4}r_\omega^{\frac{\beta-1}{2\beta}}$, respectively, which, due to the requirement from equation~\eqref{RenyiUncertainty2}, means that at {\em any} parameter $\alpha$ the lowest level saturates Tsallis entropy relation too. So, the Gaussian dependencies not only turn the Tsallis inequality into the identity but, the requirement for its applicability, equation~\eqref{Sobolev2}, is waived for them. This example again singles them out from all other probability distributions; for example, for the first excited orbital, inequality~\eqref{TsallisInequality1}, and accordingly,~\eqref{Sobolev1}, are:
\begin{equation}\label{HO_Tsallis2}
\left(\frac{2}{\alpha}\right)^{1/2}\!\!\pi^{-\frac{1+\alpha}{4\alpha}}\Gamma\!\left(\alpha+\frac{1}{2}\right)^{1/(2\alpha)}\!\!r_\omega^{\frac{1-\alpha}{2\alpha}}\geq\left(\frac{2}{\beta}\right)^{1/2}\!\!\pi^{-\frac{1+\beta}{4\beta}}\Gamma\!\left(\beta+\frac{1}{2}\right)^{1/(2\beta)}\!\!r_\omega^{\frac{\beta-1}{2\beta}}.
\end{equation}
Equation~\eqref{beta1} guarantees that this relation is dimensionally correct. Note that apart from the factor containing $r_\omega$, both sides of this expression have the same dependence on their parameters $\alpha$ or $\beta$. Accordingly, general Tsallis inequality, equation~\eqref{TsallisInequality1}, at $n=1$ is saturated at $\alpha=\beta=1$ only when its either side becomes a dimensionless $\pi^{-1/4}=0.7511\ldots$, as it was mentioned in the Introduction. Figure~\ref{HOTsallisFig1} depicts dimensionless parts of inequality~\eqref{HO_Tsallis2} (i.e., it is assumed there that $r_\omega\equiv1$) as functions of parameter $\alpha$. This manifests that for the first excited HO orbital the Tsallis inequality is satisfied at the interval defined by equation~\eqref{Sobolev2} only, in contrast to the R\'{e}nyi entropies relations, which, as mentioned above, are valid at any $\alpha\geq1/2$. For completeness, let us mention that the position and momentum parts of inequality~\eqref{HO_Tsallis2} at $\alpha=1/2$ take the values of $2/\pi^{3/4}=0.8475\ldots$ and $\left[2\!\left/\!\left(e\pi^{1/2}\right)\right.\right]^{1/2}=0.6442\ldots$ whereas at $\alpha=\infty$ they interchange these magnitudes. Also, in the vicinity of $\alpha=1$, this relation turns to
\begin{equation}\label{HO_Tsallis3}
\frac{1}{\pi^{1/4}}\left[1-\frac{\gamma-1+\ln2}{2}(\alpha-1)\right]\geq\frac{1}{\pi^{1/4}}\left[1+\frac{\gamma-1+\ln2}{2}(\alpha-1)\right],
\end{equation}
and since the term $(\gamma-1+\ln2)/2=0.1351\ldots$ is positive, the last inequality is satisfied for the Tsallis parameter approaching unity from the left, as expected.

\section{Attractive Robin wall}\label{sec_Robin}
By Robin wall we mean a 1D structure that limits the motion of the particle to a half-line, say, $x>0$, and at the confining surface, $x=0$, the following requirement is imposed on the position function $\Psi(x)$:
\begin{equation}\label{Robin1}
\left.\frac{d\Psi(x)}{dx}\right|_{x=0}=\frac{1}{\Lambda}\Psi(0).
\end{equation}
From the expression for the current density \cite{Landau1}
\begin{equation}\label{CurrentDensity1}
{\bf j}=-\frac{e\hbar}{m_p}\Im(\Psi^*{\bm\nabla}\Psi)
\end{equation}
it is elementary to check that at any {\em real} Robin length $\Lambda$ no current flows through the point $x=0$:
\begin{equation}\label{CurrentDensity2}
\left. j\right|_{x=0}=0\quad {\rm at}\quad\Im(\Lambda)=0.
\end{equation}
A remarkable property of this geometry is a fact that at the negative extrapolation parameter $\Lambda$ it has, in addition to the continuous spectrum with $E\geq0$, a single localized level with the energy \cite{Seba1,Pazma1,Bonneau1,Fulop1,Belchev1,Georgiou1,Olendski1}
\begin{equation}\label{RobinEnergy1}
E=-\frac{\hbar^2}{2m_p|\Lambda|^2},\quad\Lambda<0,
\end{equation}
with the corresponding position waveform vanishing at infinity:
\begin{equation}\label{RobinFunctionPsi1}
\Psi(x)=\left(\frac{2}{|\Lambda|}\right)^{1/2}\!\!\exp\!\left(-\frac{x}{|\Lambda|}\right),\quad x\geq0,
\end{equation}
and obeying the normalization condition
\begin{equation}\label{RobinNormalization1}
\int_0^\infty\Psi^2(x)dx=1.
\end{equation}
Its momentum counterpart \cite{Olendski1}
\begin{equation}\label{RobinFunctionPhi2}
\Phi(k)=\left(\frac{|\Lambda|}{\pi}\right)^{1/2}\frac{1}{1+i|\Lambda|k}
\end{equation}
obeys the normalization too:
\begin{equation}\label{RobinNormalization2}
\int_{-\infty}^\infty|\Phi(k)|^2dk=1.
\end{equation}
Accordingly, the corresponding densities are
\begin{subequations}\label{RobinDensities1}
\begin{eqnarray}\label{RobinDensities1_X1}
\rho(x)=\frac{2}{|\Lambda|}\exp\!\left(-2\frac{x}{|\Lambda|}\right)\\
\label{RobinDensities1_K1}
\gamma(k)=\frac{|\Lambda|}{\pi}\frac{1}{1+|\Lambda|^2k^2}.
\end{eqnarray}
\end{subequations}

Compared to the other two structures considered here, the attractive Robin wall is very peculiar, since for it a standard Heisenberg uncertainty suffers the same shortcomings as the Neumann well discussed in the Introduction, while the Shannon entropy inequality, equation~\eqref{ShannonInequality}, remains true \cite{Olendski1}. This is further evidence of the fact that quantum-information entropies present a more general  base for defining 'uncertainty' than standard deviations $\Delta x$ and $\Delta k$. As it is shown below, the R\'{e}nyi inequality that generalizes its Shannon counterpart also holds true for any $\alpha$ saturating at $\alpha=1/2$. The same statement is true for the Tsallis inequality, equation~\eqref{TsallisInequality1}, too.

R\'{e}nyi entropies of this single orbital read:
\begin{subequations}\label{RobinRenyi1}
\begin{eqnarray}\label{RobinRenyi1_X1}
R_\rho(\alpha)=\ln|\Lambda|-\ln2-\frac{\ln\alpha}{1-\alpha}\\
\label{RobinRenyi1_K1}
R_\gamma(\alpha)=-\ln|\Lambda|+\ln\pi+\frac{1}{1-\alpha}\ln\frac{\Gamma\!\left(\alpha-\frac{1}{2}\right)}{\pi^{1/2}\Gamma(\alpha)}.
\end{eqnarray}
\end{subequations}
The first thing to note from these equations is the fact that the momentum R\'{e}nyi entropy exists (or, rather, has real values) not at any arbitrary non-negative parameter $\alpha$, as was the case for the HO, but only for those indexes $\alpha$ that are greater than the threshold value that in this case is equal to one half:
\begin{equation}\label{RobinThreshold1}
\alpha_{TH}^{RW}=\frac{1}{2}.
\end{equation}
Around this point, the entropy diverges logarithmically:
\begin{equation}\label{RobinLimit1}
R_\gamma(\alpha)=-2\ln\!\left(\alpha-\frac{1}{2}\right)-\ln|\Lambda|-\ln\pi+\ldots,\quad\alpha\rightarrow\frac{1}{2},
\end{equation}
as does its position counterpart near its own threshold value of zero, as follows from equation~\eqref{RobinRenyi1_X1}. In the opposite limit of the huge R\'{e}nyi parameter, $\alpha\rightarrow\infty$, the entropies are:
\begin{subequations}\label{RobinLimitInfinite1}
\begin{eqnarray}\label{RobinLimitInfinite1_X1}
R_\rho(\alpha)=\ln|\Lambda|-\ln2+\frac{\ln\alpha}{\alpha}+\ldots\\
\label{RobinLimitInfinite1_K1}
R_\gamma(\alpha)=-\ln|\Lambda|+\ln\pi+\frac{1}{2}\frac{\ln\alpha}{\alpha}+\ldots,
\end{eqnarray}
\end{subequations}
and in the neighborhood of unity, $\alpha\rightarrow1$, one has:
\begin{subequations}\label{RobinLimitShannon1}
\begin{eqnarray}\label{RobinLimitShannon1_X1}
R_\rho(\alpha)=\ln|\Lambda|-\ln2+1-\frac{1}{2}(\alpha-1)+\frac{1}{3}(\alpha-1)^2+\ldots\\
\label{RobinLimitShannon1_K1}
R_\gamma(\alpha)=-\ln|\Lambda|+\ln\pi+\ln4-\frac{\pi^2}{6}(\alpha-1)+2\zeta(3)(\alpha-1)^2+\ldots,
\end{eqnarray}
\end{subequations}
which obviously means that the R\'{e}nyi entropies just approach their Shannon counterparts \cite{Olendski1}. Next, it directly follows from equations~\eqref{RobinRenyi1} that in equations~\eqref{RenyiUncertainty1} or \eqref{RenyiUncertainty3} the sum of the two entropies is, similar to the HO, a scale-independent dimensionless quantity. Accordingly, below in this section, during the discussion of the R\'{e}nyi entropies the distances will be measured in units of $|\Lambda|$, which is a characteristic length of this system. 

\begin{figure}
\centering
\includegraphics[width=\columnwidth]{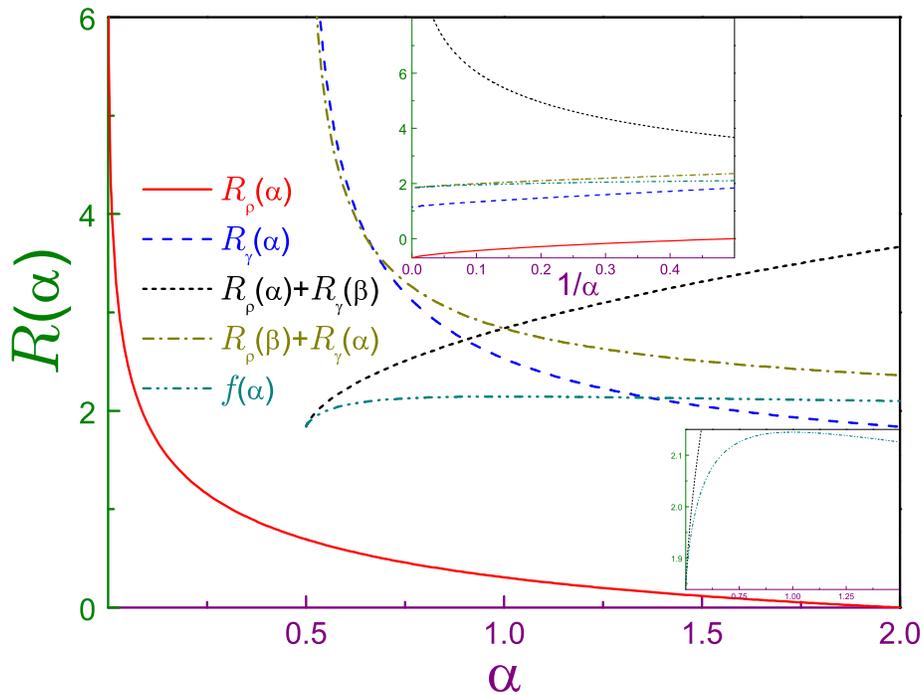}
\caption{\label{RenyiRobinFig1}
Position $R_\rho(\alpha)$ (solid line) and momentum $R_\gamma(\alpha)$ (dashed curve) R\'{e}nyi entropies of the attractive Robin wall as functions of parameter $\alpha$. Dotted line depicts $R_\rho(\alpha)+R_\gamma(\beta)$ with $\beta$ from equation~\eqref{beta1} and dash-dotted dependence is for $R_\rho(\beta)+R_\gamma(\alpha)$ with dash-dot-dotted curve representing function $f(\alpha)$ from equation~\eqref{Function_f1}. Upper inset shows the dependencies at the large R\'{e}nyi parameter (in terms of $1/\alpha$) and the lower one enlarges the view in the vicinity of $\alpha_{TH}$.}
\end{figure}

Position and momentum R\'{e}nyi entropies are shown in figure~\ref{RenyiRobinFig1} as functions of the parameter $\alpha$. As discussed above, the position component is defined on the whole non-negative axis decreasing from the infinitely high values at the infinitely small $\alpha$, passing through zero at $\alpha=2$, $R_\rho(2)=0$, and approaching $-\ln2=-0.6931\ldots$ at the parameter tending to infinity. Its momentum counterpart stays positive at any variable that is greater than its threshold monotonically decreasing from unrestrictedly high values at $\alpha=\alpha_{TH}$ to $\ln\pi=1.1447\ldots$ at the very large $\alpha$.

Turning to the discussion of the entropic uncertainty relation, we first provide asymptotic limits of the sum from the left-hand sides of equations~\eqref{RenyiUncertainty1} or \eqref{RenyiUncertainty3}:
\begin{subequations}\label{RobinLimitUncertainty1}
\begin{align}\label{RobinLimitUncertainty1_OneHalf}
R_\rho(\alpha)+R_\gamma(\beta)&=\ln2\pi+2[\ln\!\left(2\pi^{1/2}\right)-\ln(2\alpha-1)](2\alpha-1)+\ldots,\quad\alpha\rightarrow\frac{1}{2}\\
R_\rho(\alpha)+R_\gamma(\beta)&=1+\ln\pi+\ln2+\left(\frac{\pi^2}{6}-\frac{1}{2}\right)(\alpha-1)\nonumber\\
\label{RobinLimitUncertainty1_One}
&+\left[\frac{1}{3}+2\zeta(3)-\frac{\pi^2}{3}\right](\alpha-1)^2+\ldots,\,\alpha\rightarrow1\\
\label{RobinLimitUncertainty1_Infinity}
R_\rho(\alpha)+R_\gamma(\beta)&=2\ln\alpha+\ln\frac{8}{\pi}+\frac{\ln(8/\pi)-1+2\ln\alpha}{\alpha}+\ldots,\,\alpha\rightarrow\infty.
\end{align}
\end{subequations}
As the dotted line in figure~\ref{RenyiRobinFig1} shows, the sum $R_\rho(\alpha)+R_\gamma(\beta)$ is a monotonically increasing function of the parameter $\alpha$ logarithmically diverging at $\alpha$ tending to infinity, according to equation~\eqref{RobinLimitUncertainty1_Infinity}. Entropic uncertainty relation, equation~\eqref{RenyiUncertainty1} or \eqref{RenyiUncertainty3}, is always satisfied with its saturation taking place at the threshold of the momentum R\'{e}nyi entropy, as a comparison between equations~\eqref{HO_FlimitsOneHalf} and \eqref{RobinLimitUncertainty1_OneHalf} demonstrates. The behavior near the threshold is shown in enlarged format in the lower inset of the figure. For comparison, the sum $R_\rho(\beta)+R_\gamma(\alpha)$ is also depicted in figure~\ref{RenyiRobinFig1}. Its dependence is opposite to the one just described, which is the result of inversion from equation~\eqref{AlphaTransform1}; namely, its divergence at $\alpha_{TH}$ after its monotonic decrease turns at $\alpha=\infty$ into $\ln2\pi$, thus transforming the entropic uncertainty relation into the equality.

\begin{figure}
\centering
\includegraphics[width=\columnwidth]{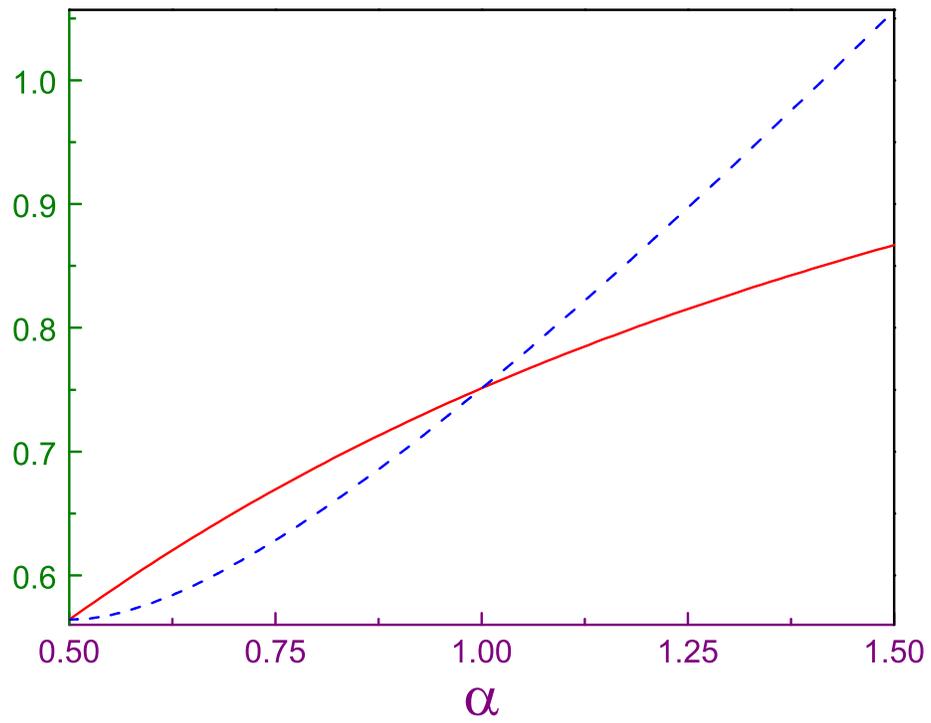}
\caption{\label{RobinTsallisFig1}
Left- (solid line) and right- (dashed curve) hand sides of relation~\eqref{RobinTsallisInequality1} as functions of the parameter $\alpha$ where it is assumed that $|\Lambda|\equiv1$.}
\end{figure}

Tsallis entropies read:
\begin{subequations}\label{RobinTsallis1}
\begin{eqnarray}\label{RobinTsallis1_X1}
T_\rho(\alpha)=\frac{1}{\alpha-1}\left[1-\frac{1}{\alpha}\left(\frac{2}{|\Lambda|}\right)^{\alpha-1}\right]\\
\label{RobinTsallis1_K1}
T_\gamma(\alpha)=\frac{1}{\alpha-1}\left[1-\frac{|\Lambda|^{\alpha-1}}{\pi^{\alpha-1/2}}\frac{\Gamma\!\left(\alpha-\frac{1}{2}\right)}{\Gamma(\alpha)}\right],
\end{eqnarray}
\end{subequations}
with their Shannon limits, $\alpha\rightarrow1$, being achieved as:
\begin{subequations}\label{RobinTsallis1a}
\begin{eqnarray}\label{RobinTsallis1a_X1}
T_\rho(\alpha)=&\!\ln|\Lambda|\!-\!\ln2\!+\!1\!-\!\left[\!-\frac{1}{2}\ln^2\!\left(\frac{2}{|\Lambda|}\right)\!+\!\ln\!\left(\frac{2}{|\Lambda|}\right)\!-\!1\!\right]\!(\alpha\!-\!1)\!+\!\ldots\\
T_\gamma(\alpha)=&-\ln|\Lambda|+\ln\pi+\ln4\nonumber\\
&+\left[-\frac{\pi^2}{6}-2\ln^2(2)-\frac{1}{2}\ln^2(\pi)-2\ln(\pi)\ln(2)\right.\nonumber\\
\label{RobinTsallis1a_K1}
&\left.-\frac{1}{2}\ln^2|\Lambda|+\ln(4\pi)\ln|\Lambda|\right](\alpha-1)+\ldots,
\end{eqnarray}
\end{subequations}
which again is different from the R\'{e}nyi asymptotes, equations~\eqref{RobinLimitShannon1}. The corresponding uncertainty relation becomes:
\begin{equation}\label{RobinTsallisInequality1}
\frac{1}{(\pi\alpha)^{1/(4\alpha)}}\left(\frac{2}{|\Lambda|}\right)^\frac{\alpha-1}{2\alpha}\geq\frac{|\Lambda|^\frac{\beta-1}{2\beta}}{\pi^{1/2}}\beta^{1/(4\beta)}\left[\frac{\Gamma\left(\beta-\frac{1}{2}\right)}{\Gamma(\beta)}\right]^{1/(2\beta)},
\end{equation}
which once more is dimensionally correct. Figure~\ref{RobinTsallisFig1} shows the dimensionless parts of this inequality. Note that this relation turns into the equality not only at $\alpha=1$, but at the R\'{e}nyi parameter being one half when its dimensionless parts, $|\Lambda|\equiv1$, degenerate to $\pi^{-1/2}=0.5641\ldots$. Near these points, inequality~\eqref{RobinTsallisInequality1} simplifies to
\begin{subequations}\label{RobinTsallis2}
\begin{eqnarray}\label{RobinTsallis2_1}
\frac{1}{\pi^{1/4}}\left[1+\frac{\ln4\pi-1}{4}(\alpha-1)\right]\geq\frac{1}{\pi^{1/4}}\left[1+\frac{\ln16\pi-1}{4}(\alpha-1)\right],\alpha\rightarrow1\\
\label{RobinTsallis2_2}
\frac{1}{\pi^{1/2}}\left[1+(\ln2\pi-1)\left(\alpha-\frac{1}{2}\right)\right]\geq\frac{1}{\pi^{1/2}}\left[1+3\left(\alpha-\frac{1}{2}\right)^2\right],\alpha\rightarrow\frac{1}{2},
\end{eqnarray}
\end{subequations}
which, obviously, are satisfied for the interval from equation~\eqref{Sobolev2} only whereas the R\'{e}nyi relation is valid for the whole region $\alpha\geq1/2$. Note the different lowest powers of the expansion coefficient in inequality~\eqref{RobinTsallis2_2}.

\section{Hydrogen atom}\label{Sec_Q1DHA}
\begin{figure}
\centering
\includegraphics[width=\columnwidth]{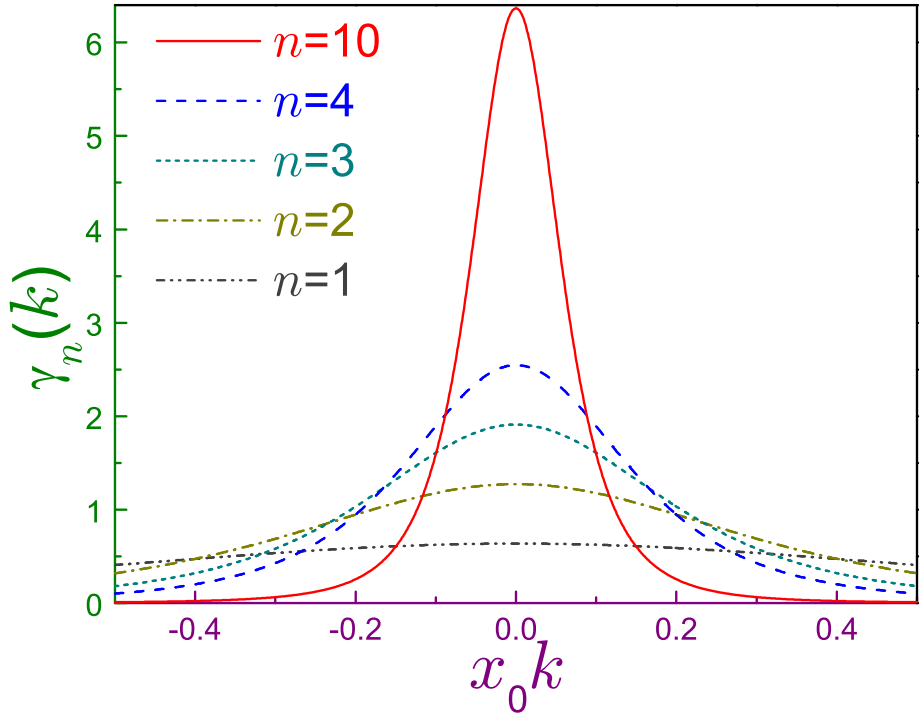}
\caption{\label{Q1DMomDensFig1}
Densities $\gamma_n(k)$ (in units of $x_0$), equation \eqref{DensityK2}, as functions of the normalized wave vector $x_0k$ where dash-dot-dotted line is for $n=1$, dash-dotted curve depicts the state with $n=2$, dotted line stands for the state with $n=3$, dashed curve -- for $n=4$, and solid line -- for $n=10$.}
\end{figure}

1D potential of the form
\begin{equation}\label{Q1Dpotential1}
V(x)=\left\{\begin{array}{cc}
-\frac{\lambda}{x},&x>0\\
\infty,&x\leq0,
\end{array}\right.
\end{equation}
$\lambda>0$, which  enters equation~\eqref{Schrodinger1}, is used, for example, for the analysis of Rydberg atoms irradiated by short half-cycle pulses \cite{Bersons1} and description of the electrons trapped above a micrometer-thick film of liquid helium \cite{P1atzman1}. The energy spectrum coincides with the 3D hydrogen atom \cite{Landau1}
\begin{equation}\label{Q1DEnergies1}
E_n=-\frac{m_p\lambda^2}{2\hbar^2n^2},\quad n=1,2,\ldots,
\end{equation}
whereas the corresponding waveforms are \cite{Omiste1,Olendski2}:
\begin{equation}\label{Q1DSolutionX1}
\Psi_n(x)=\frac{1}{x_0^{1/2}}\frac{2\overline{x}}{n^{5/2}}e^{-\overline{x}/n}L_{n-1}^{(1)}\!\left(\frac{2\overline{x}}{n}\right),
\end{equation}
with $L_n^{(\beta)}(x)$ being a generalized Laguerre polynomial \cite{Abramowitz1}. Here, we introduced a characteristic (Coulomb) length of the atom
\begin{equation}\label{Q1D_X0}
x_0=\frac{\hbar^2}{m_p\lambda},
\end{equation}
and the line over the symbol means that it is a dimensionless variable:
\begin{equation}\label{Q1DOverlineX1}
\overline{x}=\frac{x}{x_0}.
\end{equation}
Normalizable wave vector function reads \cite{Omiste1,Olendski2}:
\begin{equation}\label{SolutionP1}
\Phi_n(k)=(-1)^{n+1}x_0^{1/2}\sqrt{\frac{2n}{\pi}}\frac{\left(1-in\overline{k}\right)^{n-1}}{\left(1+in\overline{k}\right)^{n+1}},
\end{equation}
where the dimensionless wave vector is
\begin{equation}\label{Q1DOverlineK1}
\overline{k}=x_0k.
\end{equation}
Corresponding densities are
\begin{equation}\label{DensityK2}
\gamma_n(k)=\frac{2n}{\pi}\frac{x_0}{\left[1+\left(n\overline{k}\right)^2\right]^2}.
\end{equation}
Some of them are depicted in figure~\ref{Q1DMomDensFig1}, which shows the sharpening of $\gamma_n(k)$ around $k=0$ with the quantum number $n$ increasing. Didactically, it is important to note that at infinitely large index $n$ the distribution from equation~\eqref{DensityK2} turns into the $\delta$-function:
\begin{equation}\label{Q1DLimit1}
\gamma_n(k)\rightarrow\delta(k),\quad n\rightarrow\infty,
\end{equation}
which means that the quantum particle can have a zero momentum only. On the other hand, this statement is corroborated by equation~\eqref{Q1DEnergies1}, which confirms that in this quasi-classical regime the energy, and accordingly, the momentum $p_n=\sqrt{2m_pE_n}$, are zeros. Simultaneously, as follows from equation~\eqref{Q1DSolutionX1}, the position waveform for the index $n$ growing is getting flatter and approaches zero. Such opposite behavior of the original function and its image, equations~\eqref{Fourier1}, is a very general property of Fourier transformation \cite{Arfken1}.

Observe that the expression for the 1D position waveform, equation~\eqref{Q1DSolutionX1}, is very similar to the angle-independent, $l=0$, function of the 3D hydrogen-like atom \cite{Messiah1}:
\begin{equation}\label{3DHydrogenPosition1}
\Psi_{nlm}^{3D}\!\left(r,\theta_r,\varphi_r\right)=\frac{1}{x_0^{3/2}}\frac{2}{n^2}\sqrt{\frac{(n-l-1)!}{(n+l)!}}\!\left(\frac{2\overline{r}}{n}\right)^l\!\!e^{-\overline{r}/n}\!L_{n-l-1}^{2l+1}\!\left(\frac{2\overline{r}}{n}\right)\!Y_l^m\left(\theta_r,\varphi_r\right),
\end{equation}
written in the spherical coordinates ${\bf r}\equiv(r,\theta_r,\varphi_r)$\footnote{Due to the several different definitions of the Laguerre polynomials, different expressions are used for the function $\Psi_{nlm}^{3D}\!\left(r,\theta_r,\varphi_r\right)$; for example, Landau and Lifshitz \cite{Landau1} write $L_{n+l}^{2l+1}$ instead of our $L_{n-l-1}^{2l+1}$ since their terminology is different from Abramowitz and Stegun \cite{Abramowitz1} and Bateman \cite{Bateman1} to which we adhere here (cf. equation~(d.13) in \cite{Landau1} versus equations~(10.12.5) and (10.12.7) in \cite{Bateman1}). Also, Messiah's definition, equation~(B.13) in \cite{Messiah1}, is slightly different from ours too.}. Here, $n=1,2,\ldots$ is a principal quantum number, $l=0,1,\ldots,n-1$ is an azimuthal quantum number, and $m=-l,-l+1\ldots,l-1,l$ is a magnetic index, and $Y_l^m(\theta,\varphi)$ is a standard spherical harmonics. Saha, Talukdar and Chatterjee (STC) \cite{Saha1}, inspired by the formal identity of 1D equation~\eqref{Q1DSolutionX1} and its spherically symmetric, $l=0$, 3D counterpart from equation~\eqref{3DHydrogenPosition1}\footnote{Actually, for $l=0$ there is no in the resulting expression the term corresponding to the factor $\overline{x}$ in equation~\eqref{Q1DSolutionX1} but, probably, it went unnoticed by STC.}, claim that since the 1D and 3D Fourier originals are the same, their Fourier images should be identical too. Accordingly, instead of using a direct Fourier transform
\begin{equation}\label{Fourier2}
\Phi_n(k)=\frac{1}{\sqrt{2\pi}}\int_0^\infty\Psi_n(x)e^{-ikx}dx,
\end{equation}
as follows from equation~\eqref{Fourier1_1} and was done in references \cite{Olendski2,Omiste1}, they take a general expression of the 3D momentum function derived soon after the birth of quantum mechanics \cite{Podolsky1}:
\begin{eqnarray}
\Phi_{nlm}^{3D}\!\left(k,\theta_k,\varphi_k\right)&=x_0^{3/2}\sqrt{\frac{2}{\pi}\frac{(n-l-1)!}{(n+l)!}}2^{2(l+1)}n^2l!\frac{\left(n\overline{k}\right)^l}{\left[1+\left(n\overline{k}\right)^2\right]^{l+2}}\nonumber\\
\label{3DHydrogenMomentum1}
&\times C_{n-l-1}^{l+1}\!\left(\frac{\left(n\overline{k}\right)^2-1}{\left(n\overline{k}\right)^2+1}\right)\!Y_l^m\left(\theta_k,\varphi_k\right)
\end{eqnarray}
(where $C_n^\lambda(x)$ is a Gegenbauer polynomial \cite{Abramowitz1,Bateman1}), zero in it the azimuthal quantum number, $l=0$, (and, accordingly, magnetic index too, $m=0$) and announce that the ultimate expression containing Chebyshev polynomial of the second kind $U_n(x)$ \cite{Abramowitz1,Bateman1} is a correct representation of the momentum waveform of the Q1D hydrogen atom. In addition, they claim that since $r$ in equation~\eqref{3DHydrogenPosition1} with $l=m=0$ changes from zero to infinity, the same holds true in their expression for the momentum component too, $0\leq k<\infty$, thus arbitrarily forbidding motion of the particle to the left. To refute this faulty analysis, let us point out first that even though equations~\eqref{Q1DSolutionX1} and \eqref{3DHydrogenPosition1}  with $l=0$ look formally the same (let us forget now about the missing multiplier corresponding to $\overline{x}$), they describe completely different geometries: for the Q1D case, the particle motion is strictly limited to the positive half-line and for the 3D configuration the length $r$ is an absolute value of the radius vector $\bf r$, which by definition can not be negative, whereas each Cartesian coordinate
\begin{eqnarray}
x&=r\sin\theta_r\cos\varphi_r\nonumber\\
y&=r\sin\theta_r\sin\varphi_r\nonumber\\
z&=r\cos\theta_r\nonumber
\end{eqnarray}
varies along the whole axis, $-\infty<x,y,z<+\infty$. Then, for finding its Fourier transform, one uses
\begin{eqnarray}
\Phi_{nlm}^{3D}=\int_{-\infty}^{+\infty}dx\int_{-\infty}^{+\infty}dy\int_{-\infty}^{+\infty}dz\Psi^{3D}e^{-i(k_xx+k_yy+k_zz)}\nonumber\\
=\int_0^{+\infty}dr\int_0^\pi d\theta_r\int_0^{2\pi}d\varphi_rr^2\sin\theta_r\Psi_{nlm}^{3D}(r,\theta_r,\varphi_r)\nonumber\\
\label{3DHydrogenIntergation1}
\times e^{-ikr[\cos\theta_k\cos\theta_r+\sin\theta_k\sin\theta_r\cos(\varphi_r-\varphi_k)]},
\end{eqnarray}
and the last integral is calculated as it was done by Podolsky and Pauling \cite{Podolsky1} leading to equation~\eqref{3DHydrogenMomentum1}. The first equality was inserted into equation~\eqref{3DHydrogenIntergation1} in order to explicitly underline that each Cartesian component is running through the whole axis. Accordingly, $k$ in equation~\eqref{3DHydrogenMomentum1} is a length of the wave vector which is non-negative but each of its Cartesian components, due to isotropy of the problem, can take either a positive or negative axis, $-\infty<k_x,k_y,k_z<+\infty$. In the same way, in equation~\eqref{SolutionP1}, which was derived through the integration from equation~\eqref{Fourier2}, reference \cite{Olendski2}, the momentum $k$ runs over the whole of the real values, $-\infty<k<+\infty$, since the motion to the right ($k>0$) for this geometry does not have any advantages compared with the movement to the left ($k<0$). Not surprisingly, the 1D, equation~\eqref{Fourier2}, and 3D, equation~\eqref{3DHydrogenIntergation1}, integrations yield different results. The flawed nature of the STC momentum wave function can be further exemplified by calculating with its help the momentum Shannon entropy that, as mentioned below, violates the corresponding inequality, equation~\eqref{ShannonInequality}.

Position R\'{e}nyi and Tsallis entropies can be calculated analytically for the ground state only when the corresponding Laguerre polynomial is just a unity, $L_0^{(\beta)}(x)\equiv1$, whereas their momentum counterparts can be easily evaluated for any level:
\begin{subequations}\label{FunctionalsQ1D}
\begin{eqnarray}
\label{RenyiQ1D_X1}
R_{\rho_1}(\alpha)&=\ln x_0+\frac{1}{1-\alpha}\ln\frac{\Gamma(2\alpha+1)}{2\alpha^{2\alpha+1}}\\
\label{RenyiQ1D_P1}
R_{\gamma_n}(\alpha)&=-\ln x_0-\ln n+\frac{1}{1-\alpha}\ln\!\!\left(\frac{2^\alpha}{\pi^{\alpha-1/2}}\frac{\Gamma\!\left(2\alpha-\frac{1}{2}\right)}{\Gamma(2\alpha)}\right)\\
\label{TsallisQ1D_X1}
T_{\rho_1}(\alpha)&=\frac{1}{\alpha-1}\left[1-\frac{1}{x_0^{\alpha-1}}\frac{\Gamma(2\alpha+1)}{2\alpha^{2\alpha+1}}\right]\\
\label{TsallisQ1D_P1}
T_{\gamma_n}(\alpha)&=\frac{1}{\alpha-1}\left[1-x_0^{\alpha-1}\frac{\pi^{1/2}}{n}\left(\frac{2n}{\pi}\right)^\alpha\frac{\Gamma\left(2\alpha-\frac{1}{2}\right)}{\Gamma(2\alpha)}\right].
\end{eqnarray}
\end{subequations}
Note that momentum components are finite for $\alpha\geq\alpha_{TH}^{Q1D}=1/4$ only whereas position entropies have their values defined for any non-negative parameter. It is important to stress that the Q1D critical magnitude $\alpha_{TH}^{Q1D}$ is different to the one of the Robin wall $\alpha_{TH}^{RW}$, equation~\eqref{RobinThreshold1}, and HO, which is zero. Thus, the range where the {\em momentum} R\'{e}nyi and Tsallis entropies are defined strongly depends on the corresponding {\em position} potential $V(x)$ and the boundary conditions. Asymptotic limits are:

for the R\'{e}nyi entropies:
\begin{subequations}\label{LimitsQ1D_R}
\begin{eqnarray}
\label{LimitR_Q1D_X0}
R_{\rho_1}(\alpha)&=\ln x_0-\ln(2\alpha)-(2\gamma+\ln2+3\ln\alpha)\alpha+\ldots,\quad\alpha\rightarrow0\\
R_{\rho_1}(\alpha)&=\ln x_0+2\gamma+\left(3-\frac{\pi^2}{3}\right)(\alpha-1)\nonumber\\
\label{LimitR_Q1D_X1}
&+\left[\frac{8}{3}\,\zeta(3)-3\right](\alpha-1)^2+\ldots,\quad\alpha\rightarrow1\\
\label{LimitR_Q1D_Xinf}
R_{\rho_1}(\alpha)&=\ln x_0+2-\ln4+\frac{4-\ln(16\pi)+\ln\alpha}{\alpha}+\ldots,\quad\alpha\rightarrow\infty\\
\label{LimitR_Q1D_P14}
R_{\gamma_n}(\alpha)&=-\ln x_0-\frac{4}{3}\ln\!\left[\pi^{1/4}(2n)^{3/4}\left(\alpha-\frac{1}{4}\right)\right]+\ldots,\quad\alpha\rightarrow\frac{1}{4}+0\\
R_{\gamma_n}(\alpha)&=-\ln x_0-2+\ln\frac{8\pi}{n}+\left(6-\frac{2}{3}\pi^2\right)(\alpha-1)\nonumber\\
\label{LimitR_Q1D_P1}
&+\left[16\zeta(3)-\frac{56}{3}\right](\alpha-1)^2+\ldots,\quad\alpha\rightarrow1\\
\label{LimitR_Q1D_Pinf}
R_{\gamma_n}(\alpha)&=-\ln x_0+\ln\frac{\pi}{2n}+\frac{\ln\alpha+\ln\frac{\pi}{2}}{2\alpha}+\ldots,\quad\alpha\rightarrow\infty;
\end{eqnarray}
\end{subequations}

for the Tsallis entropies:
\begin{subequations}\label{LimitsQ1D_T}
\begin{eqnarray}
\label{LimitT_Q1D_X0}
T_{\rho_1}(\alpha)&=\frac{x_0}{2\alpha}-1-\frac{1}{2}(2\gamma+\ln x_0)x_0-x_0\ln\alpha+\frac{x_0}{2}+\ldots,\quad\alpha\rightarrow0\\
T_{\rho_1}(\alpha)&=\ln x_0+2\gamma+\left(3-\frac{\pi^2}{3}-2\gamma^2-2\gamma\ln x_0-\frac{1}{2}\ln^2x_0\right)(\alpha-1)\nonumber\\
&+\left[\frac{1}{6}\ln^3x_0+\gamma\ln^2x_0+\frac{1}{6}\left(2\pi^2+12\gamma^2-18\right)\ln x_0\right.\nonumber\\
\label{LimitT_Q1D_X1}
&+\left.\frac{8}{3}\,\zeta(3)+\frac{2}{3}\gamma\pi^2+\frac{4}{3}\gamma^3-6\gamma-3\right](\alpha-1)^2+\ldots,\quad\alpha\rightarrow1\\
\label{LimitT_Q1D_Xinf}
T_{\rho_1}(\alpha)&=\frac{1}{\alpha}+\frac{1}{\alpha^2}+\ldots,\quad\alpha\rightarrow\infty\\
\label{LimitT_Q1D_P14}
T_{\gamma_n}(\alpha)&=\frac{2}{3}\left(\frac{2}{\pi n^3}\right)^{1/4}\frac{1}{x_0^{3/4}}\frac{1}{\alpha-\frac{1}{4}}-\frac{4}{3}+\ldots,\quad\alpha\rightarrow\frac{1}{4}+0,
\end{eqnarray}
\end{subequations}
where in equation~\eqref{LimitT_Q1D_Xinf} it is additionally assumed that $x_0\geq1$. Since the expressions for the Tsallis momentum components at $\alpha\rightarrow1$ and $\alpha\rightarrow\infty$ are quite unwieldy, they are not shown here. It is observed once again that R\'{e}nyi and Tsallis  functionals approach their Shannon limit in different ways. Note that for the ground state the Shannon entropies are $S_{\rho_1}=\ln x_0+2\gamma$ and $S_{\gamma_1}=-\ln x_0-2+\ln(8\pi)$ with their sum $S_{\rho_1}+S_{\gamma_1}=2\gamma-2+\ln(8\pi)=2.3786\ldots$ satisfying, of course, Shannon uncertainty relation, equation~\eqref{ShannonInequality}. Contrary, the STC momentum Shannon entropy with the wave function from reference \cite{Saha1} that can be calculated only numerically violates this inequality since $S_{\gamma_1}^{STC}=-\ln x_0+0.5575\ldots$ and, accordingly, $S_{\rho_1}+S_{\gamma_1}^{STC}=1.7119\ldots$, what is another proof of its faulty nature.

\begin{figure}
\centering
\includegraphics[width=\columnwidth]{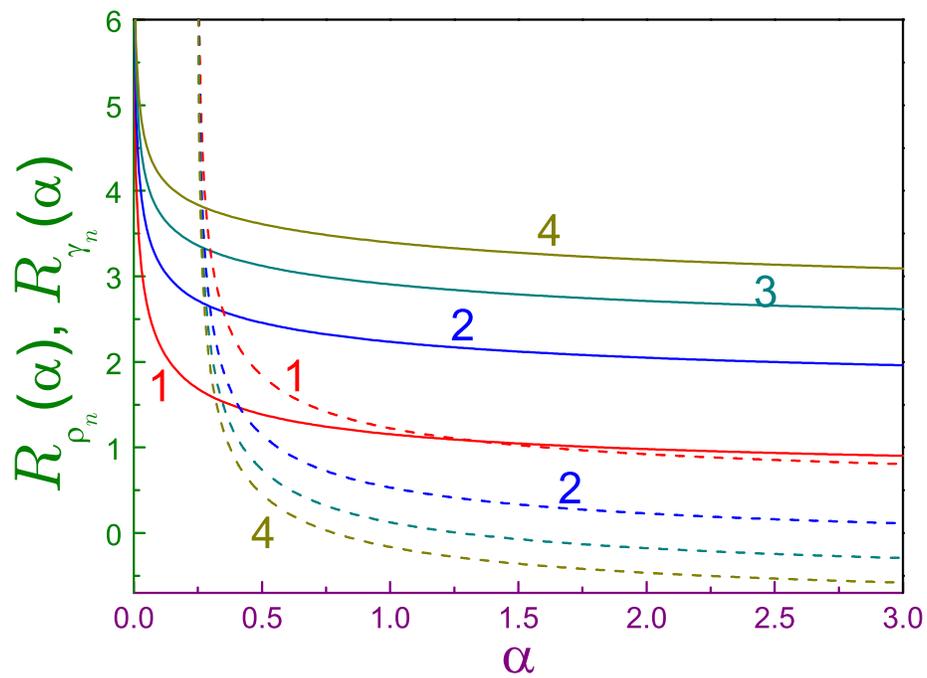}
\caption{\label{Q1DRenyiFig1}
R\'{e}nyi position $R_{\rho_n}$ (solid lines) and momentum $R_{\gamma_n}$ (dashed curves) entropies of the Q1D hydrogen atom as functions of the parameter $\alpha$. Numbers near the curves denote quantum index $n$.}
\end{figure}

R\'{e}nyi entropies of the four lowest-energy levels are shown in figure~\ref{Q1DRenyiFig1} where the distances are measured in units of $x_0$ and position parts for $n\geq2$ were evaluated numerically. As derived in equations~\eqref{LimitR_Q1D_X0} and \eqref{LimitR_Q1D_P14}, position and momentum components diverge logarithmically when the parameter $\alpha$ approaches their corresponding threshold values that are zero and $1/4$, respectively. Observe that at any R\'{e}nyi parameter the position part is an increasing function of the quantum index whereas its momentum counterpart decreases as $-\ln n$, equation~\eqref{RenyiQ1D_P1}, contrary to the HO where both $R_{\rho_n}$ and $R_{\gamma_n}$, which are equal to each other, get larger for the larger $n$, see figure~\ref{HOFig1}.

\begin{figure}
\centering
\includegraphics[width=\columnwidth]{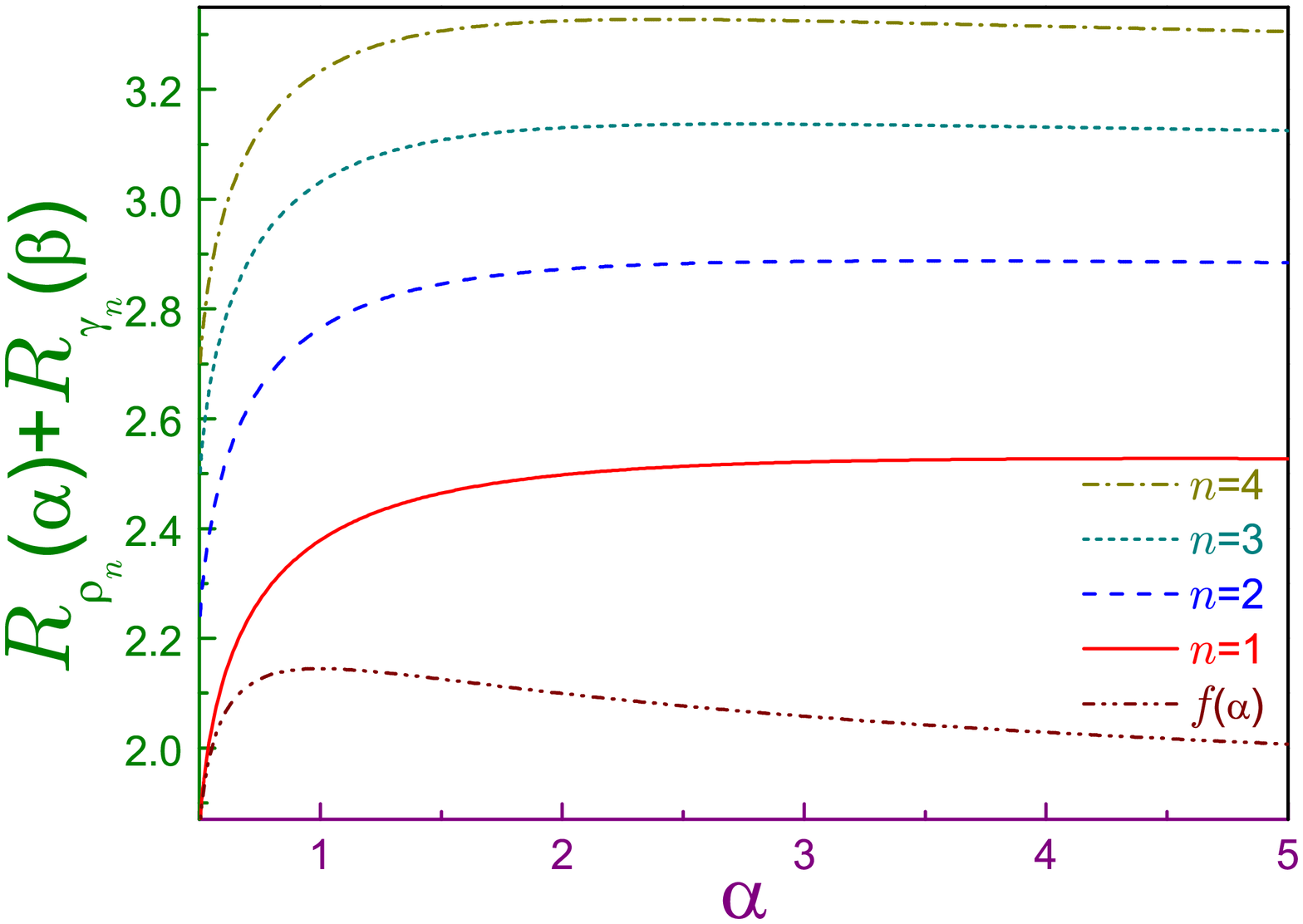}
\caption{\label{Q1DRenyiUncertaintyFig1}
Sum of the position and momentum R\'{e}nyi entropies $R_{\rho_n}(\alpha)+R_{\gamma_n}(\beta)$ with parameter $\beta$ from equation~\eqref{beta1} of the four lowest states of Q1D hydrogen atom as functions of the parameter $\alpha$. The function $f(\alpha)$ from equation~\eqref{Function_f1} is shown by dash-dot-dotted line.}
\end{figure}

For the ground orbital, the sum of the two entropies entering R\'{e}nyi relations~\eqref{RenyiUncertainty1} or \eqref{RenyiUncertainty3} has the following limits:
\begin{subequations}\label{ReyiUncertaintyQ1D1}
\begin{eqnarray}
R_{\rho_1}(\alpha)+R_{\gamma_1}(\beta)=\ln2\pi\nonumber\\
\label{ReyiUncertaintyQ1D1_LimitOneHalf}
+[\ln8-\gamma-1-\ln(2\alpha-1)](2\alpha-1)+\ldots,\quad\alpha\rightarrow\frac{1}{2}\\
R_{\rho_1}(\alpha)+R_{\gamma_1}(\beta)=-2+2\gamma+\ln8\pi+\left(\frac{\pi^2}{3}-3\right)(\alpha-1)\nonumber\\
\label{ReyiUncertaintyQ1D1_LimitOne}
+\left[-\frac{4}{3}\pi^2+\frac{56}{3}\zeta(3)-\frac{29}{3}\right](\alpha-1)^2+\ldots,\quad\alpha\rightarrow1\\
\label{ReyiUncertaintyQ1D1_LimitInfinity}
R_{\rho_1}(\alpha)+R_{\gamma_1}(\beta)=2+\ln\frac{\pi}{2}+\frac{2-\ln8-\frac{1}{2}\ln\pi+\frac{1}{2}\ln\alpha}{\alpha},\quad\alpha\rightarrow\infty,
\end{eqnarray}
\end{subequations}
with $-2+2\gamma+\ln8\pi=2.3786\ldots$ and $2+\ln\frac{\pi}{2}=2.4515\ldots$. Comparing equation~\eqref{ReyiUncertaintyQ1D1_LimitOneHalf} with its counterpart for the function $f(\alpha)$, equation~\eqref{HO_FlimitsOneHalf}, one sees that the limit $\alpha\rightarrow1/2$ turns the R\'{e}nyi uncertainty relation into the identity and, since $\ln8-\gamma-1=0.5022\ldots$ is positive, equation~\eqref{RenyiUncertainty1} in the vicinity of this asymptote holds true. Figure~\ref{Q1DRenyiUncertaintyFig1} depicts the sum $R_{\rho_n}(\alpha)+R_{\gamma_n}(\beta)$ for the four lowest-energy states together with the function $f(\alpha)$. As just discussed, for the ground orbital the point $\alpha=1/2$ saturates the R\'{e}nyi inequality. Numerical analysis shows the $R_{\rho_1}(\alpha)+R_{\gamma_1}(\beta)$ has a very broad maximum of $2.5273$ located at $\alpha\approx4.55$. Strictly speaking, one can find this extremum by zeroing a derivative with respect to $\alpha$ of the corresponding sum but the resulting equation has a quite complicated form and its analytic solution is impossible. Figure also exemplifies that the sum of the two entropies is an increasing function of the quantum index, $R_{\rho_{n+1}}(\alpha)+R_{\gamma_{n+1}}(\beta)>R_{\rho_n}(\alpha)+R_{\gamma_n}(\beta)$. Let us also point out that for the larger $n$ its maximum shifts to the smaller R\'{e}nyi parameter; for example, for the fist excited level the extremum of $2.8876\ldots$ is achieved at $\alpha\approx3.53$, and for next two orbitals these numbers are: $3.1370\ldots$, $\alpha\approx2.77$, and $3.3277\ldots$, $\alpha\approx2.42$, respectively.

\begin{figure}
\centering
\includegraphics[width=0.9\columnwidth]{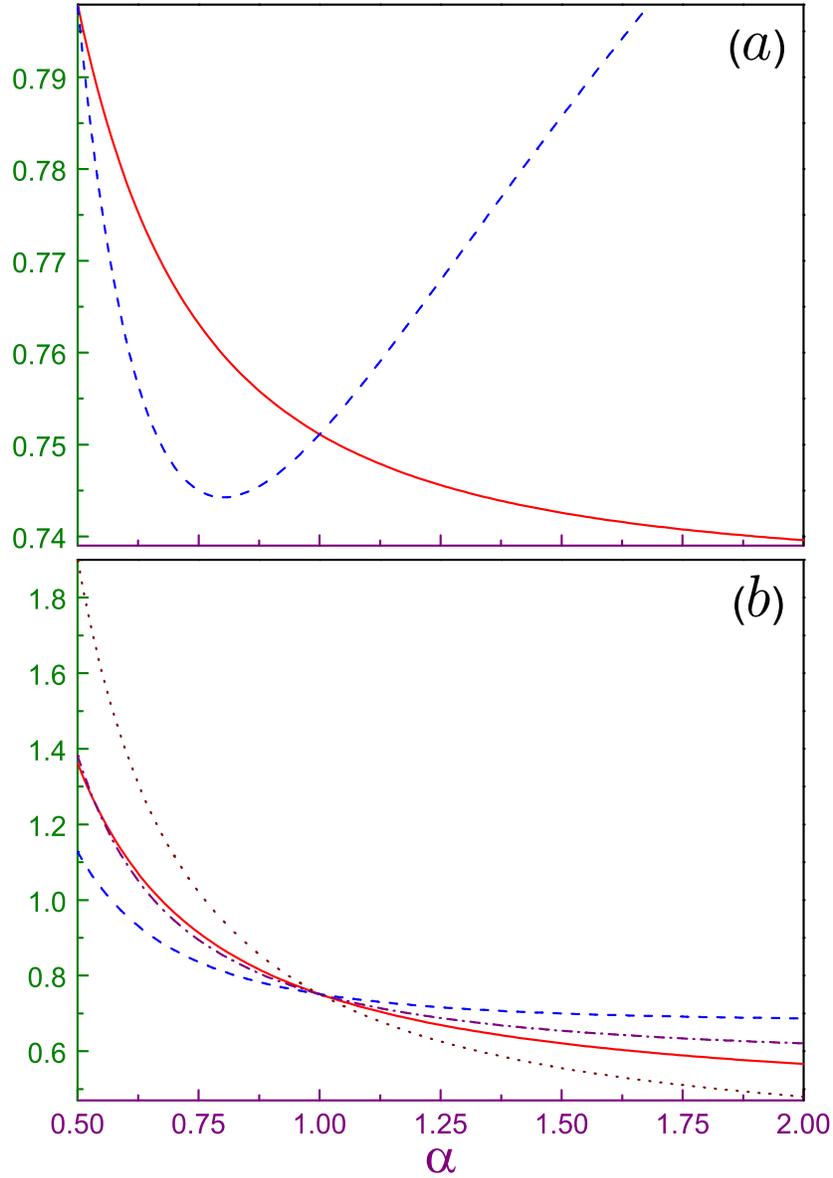}
\caption{\label{Q1DTsallisUncertaintyFig1}
(a) Left- (solid line) and right- (dashed curve) hand sides of relation~\eqref{TsallisInequalityQ1D1} as functions of the parameter $\alpha$ where it is assumed that $x_0\equiv1$. (b) Solid (dashed) line presents a left (right) dimensionless part of Tsallis uncertainty relation, equation~\eqref{TsallisInequality1}, for the first excited state of the Q1D hydrogen atom whereas the dotted (dash-dotted) curve depicts the same quantity for the second excited level.}
\end{figure}

Tsallis inequality, equation~\eqref{TsallisInequality1}, for the lowest bound level degenerates to
\begin{equation}\label{TsallisInequalityQ1D1}
x_0^{(1-\alpha)/(2\alpha)}\frac{\Gamma(2\alpha+1)^{1/(2\alpha)}}{\pi^{1/(4\alpha)}2^{1/(2\alpha)}\alpha^{1+1/(4\alpha)}}\geq x_0^{(\beta-1)/(2\beta)}\!\left(\frac{2}{\pi}\right)^{1/2}\!\left[\frac{\Gamma\!\left(2\beta-\frac{1}{2}\right)}{\Gamma(2\beta)}\right]^{1/(2\beta)}\!\!\!\!\!\!\!\!\!.
\end{equation}
As before, dimension of the left side matches that of the right. Accordingly, in our analysis below we drop the term with $x_0$. In the limiting cases this relation simplifies to:
\begin{subequations}\label{TsallisInequalityQ1D2}
\begin{eqnarray}\label{TsallisInequalityQ1D2_1}
\sqrt{\frac{2}{\pi}}\left[\!1\!-\!(\ln2)\!\left(\!\alpha\!-\!\frac{1}{2}\!\right)\!\right]\geq\sqrt{\frac{2}{\pi}}\!\left[\!1\!-\!(2\gamma\!+\!1\!-\!\ln(2\pi))\!\left(\!\alpha\!-\!\frac{1}{2}\!\right)\!\right],\,\alpha\rightarrow\frac{1}{2},\\
\label{TsallisInequalityQ1D2_2}
\frac{1}{\pi^{1/4}}\!\!\left[\!1\!+\!\frac{1-4\gamma+\ln\pi}{4}(\!\alpha\!-\!1)\right]\geq\frac{1}{\pi^{1/4}}\!\!\left[\!1\!+\!\frac{\ln(64\pi)-5}{4}(\!\alpha\!-\!1)\right],\,\alpha\rightarrow1,
\end{eqnarray}
\end{subequations}
where $(2/\pi)^{1/2}=0.7978\ldots$. Note that since $2\gamma+1-\ln(2\pi)=0.3165\ldots$ is smaller than $\ln2=0.6931\ldots$ and since $(1-4\gamma+\ln\pi)/4=-0.04103\ldots$ is negative whereas $[\ln(64\pi)-5]/4=0.07590\ldots$ is positive, equations~\eqref{TsallisInequalityQ1D2} prove that Tsallis uncertainty relation holds only inside the interval from equation~\eqref{Sobolev2}. Behaviour of the dimensionless parts from relation~\eqref{TsallisInequalityQ1D1} is shown in panel (a) of figure~\ref{Q1DTsallisUncertaintyFig1}. It is seen that, as it follows from equation~\eqref{TsallisInequalityQ1D2_1}, saturation of the corresponding relation takes place not only at $\alpha=1$, equation~\eqref{TsallisInequalityQ1D2_2}, but at $\alpha=1/2$ too, as it was the case for the Robin wall, Chapter~\ref{sec_Robin}. However, this property is not a universal characteristics of the Q1D hydrogen atom since for the higher-lying states the only saturation point is just the right edge of this interval. This is exemplified in figure~\ref{Q1DTsallisUncertaintyFig1}(b) that depicts both sides of the Tsallis inequality~\eqref{TsallisInequality1} for the first two excited levels. In fact, as our numerical results disclose, the difference between the position and momentum parts of equation~\eqref{TsallisInequality1} at $\alpha=1/2$ increases with the quantum index $n$.

\section{Conclusions}
One-parameter quantum-information measures find more and more applications in different branches of science and other spheres of human activity. We provided in Chapter~\ref{Sec_Intro} a by no means complete list of the fields where the R\'{e}nyi $R(\alpha)$ and Tsallis $T(\alpha)$ entropies have been used. There is no doubt that this catalog will expand. To introduce to the students the basic concepts of these two functionals, we considered here three 1D quantum systems that allow (fully or at least partially) an analytic calculation of the position and momentum components of these measures; namely, position Schr\"{o}dinger equation~\eqref{Schrodinger1} yields solutions $\Psi(x)$ in terms of known to the undergraduate students functions, such as Hermite or Laguerre polynomials, and subsequent calculation of their Fourier transforms according to equation~\eqref{Fourier1_1} leads to the momentum waveforms $\Phi(k)$, which again are expressed analytically. After this, the mathematical road to the entropies lies through the integration according to equations~\eqref{Functionals1}. In this way, during the consideration of the HO, Section~\ref{sec_HO}, a special cachet of the Gaussian distribution \cite{DePalma1}, which corresponds to the lowest-energy state of this geometry, was reconfirmed: at any dimensionless parameter $\alpha$ it saturates not only the R\'{e}nyi uncertainty relation, equation~\eqref{RenyiUncertainty1}, but also its Tsallis counterpart too, equation~\eqref{TsallisInequality1}, without taking into account a restriction from equation~\eqref{Sobolev2}, which is crucial for any other probability arrangement. Right-hand side of the R\'{e}nyi inequality~\eqref{RenyiUncertainty1}, which determines the limit of the simultaneous knowledge of the position and momentum, has a maximum at $\alpha=1$ which physically means, according to our interpretation provided in the Introduction, that just the Shannon entropy among all other $\alpha$ provides less total information about particle location and motion. Higher-energy HO orbitals obey the rule: $R_{n+1}(\alpha)>R_n(\alpha)$ and the sum from the uncertainty relation is an increasing function of the quantum index too with its maximum for the arbitrary $n$ being achieved at the Shannon case, $\alpha=1$. Interestingly, for the Q1D hydoren-like atom the position R\'{e}nyi entropies, similar to the HO, increase with $n$ whereas momentum components move in the opposite direction as $-\ln n$. Importantly, for all three geometries the obtained results simplify in the limiting values of parameter $\alpha$; for example, comparing the entropies behaviour near the Shannon case, $\alpha\rightarrow1$, we have shown that even though at $\alpha=1$ both R\'{e}nyi $R$ and Tsallis $T$ functionals reduce to the Shannon counterparts, the trajectories along which this transformation takes place, for all three configurations are different for $R$ and $T$ in either space. Each system exhibits its own behaviour of the sum $R_\rho(\alpha)+R_\gamma(\beta)$ that enters the uncertainty relation from equation~\eqref{RenyiUncertainty1}: for the HO, it has an extremum at $\alpha=1$ for all levels whereas the $n$-dependent maximum of the hydrogenic orbitals is shifted on the $\alpha$ axis to the left  with the energy growing, and for the Robin wall it monotonically and unrestrictedly increases from its saturation value at the R\'{e}nyi parameter one half. Next important conclusion lies in the fact that depending on the form of the \emph{position} potential $V(\bf r)$ in the Shcr\"{o}dinger equation or boundary conditions imposed on the \emph{position} function $\Psi({\bf r})$, the semi-infinite range of the R\'{e}nyi or Tsallis parameter where the \emph{momentum} component exists is different with its lowest edge being zero for the HO, one half for the attractive Robin wall and one quarter for the hydrogen structure. This topic of the influence of the \emph{position} parameters on the properties of the \emph{momentum} components needs further development and analysis. Also, it follows from the above consideration that at $\alpha=1/2$ the lowest orbital of all three structures transforms R\'{e}nyi and Tsallis uncertainty relations into the equalities. To check whether or not this is a general feature of any quantum object, we have performed corresponding calculations for the zero-energy state of the 1D Neumann well mentioned in Sec.~\ref{Sec_Intro}: for it, the left-hand side of equation~\eqref{RenyiUncertainty1} takes the form
\begin{equation}\tag{85a}\label{eq:85a}
R_{\rho_0}^N(\alpha)+R_{\gamma_0}^N(\beta)=\frac{1}{1-\beta}\ln\!\!\left(\left(\frac{2}{\beta}\right)^\beta\int_{-\infty}^\infty\left[\left(\frac{\sin\frac{z}{2}}{z}\right)^2\right]^\beta dz\right),
\end{equation}
and, as our numerical results show, at $\alpha\rightarrow1/2$ this expression really does converge to $\ln2\pi$. At the same time, its Tsallis uncertainty relation reads:
\begin{equation}\tag{85b}\label{eq:85b}
a^\frac{1-\alpha}{2\alpha}\left(\frac{\alpha}{\pi}\right)^\frac{1}{4\alpha}\geq a^\frac{\beta-1}{2\beta}\left(\frac{2}{\pi}\right)^{1/2}\left(\frac{\beta}{\pi}\right)^\frac{1}{4\beta}\left(\int_{-\infty}^\infty\left[\left(\frac{\sin\frac{z}{2}}{z}\right)^2\right]^\beta dz\right)^\frac{1}{2\beta},
\end{equation}
which is indeed satisfied as inequalty at any Tsallis parameter smaller than unity while the asymptote $\alpha=1/2$ turns it into the identity with either side becoming $(2\pi)^{-1/2}=0.3989\ldots$. The same property persists for the Dirichlet well too (not shown here). In this way, we arrive at the following 

{\bf Conjecture:}

\emph{Lowest orbital of the $l$-dimensional quantum structure in the limit $\alpha=1/2$ transforms R\'{e}nyi, equation~\eqref{RenyiUncertainty1}, and Tsallis, equation~\eqref{TsallisInequality1},  uncertainty relations into the equalities}.

General proof of this statement (if it is true) lies beyond the scope of the present pedagogical article.
 
\section{Acknowledgments}
Research was supported by SEED Project No. 1702143045-P from the Research Funding Department, Vice Chancellor for Research and Graduate Studies, University of Sharjah.


\section*{References}
\begin{thebibliography}{100}
\bibitem{Renyi1}R\'{e}nyi A 1960 On measures of Information Theory {\em Proceedings of the Fourth Berkeley Symposium on Mathematics, Statistics and Probability} (Berkeley University Press: Berkeley, CA, USA)
\bibitem{Renyi2}R\'{e}nyi A 1970 {\em Probability Theory} (Amsterdam: North-Holland)
\bibitem{Tsallis1}Tsallis C 1988 Possible generalization of Boltzmann-Gibbs statistics\JSP{\bf 52} 479--87
\bibitem{Havrda1}Havrda J and Charv\'{a}t F 1967 Quantification method of classification processes. Concept of structural {\em a}-entropy {\em Kybernetika} {\bf 3} 30--5
\bibitem{Daroczy1}Dar\'{o}czy Z 1970 Generalized information functions {\em Inform. Control} {\bf 16} 36--51
\bibitem{Shannon1}Shannon C E 1948 A mathematical theory of communication {\em Bell Syst. Tech. J.} {\bf 27} 379--423
\bibitem{Onicescu1}Onicescu O 1966 \'{E}nergie informationnelle\CRA{\bf 263} 841--2
\bibitem{Rajagopal1}Rajagopal A K 1995 The Sobolev inequality and the Tsallis entropic uncertainty relation\PLA{\bf 205} 32--6
\bibitem{Jizba1}Jizba P and Arimitsu T 2001 The world according to R\'{e}nyi: Thermodynamics of multifractal systems {\em AIP Conf. Proc.} {\bf 597} 341--8

Jizba P and Arimitsu T 2004 The world according to R\'{e}nyi: Thermodynamics of multifractal systems\APNY{\bf  312} 17--59
\bibitem{Bashkirov1}Bashkirov A G 2006 Entropia R\'{e}nyi kak statisticheskaya entropia dlia slozhnykh system {\em Teor. Mat. Fiz.} {\bf 149} 299--317

(English transl.: 2006 R\'{e}nyi entropy as a statistical entropy for complex systems {\em Theor. Math. Phys.} {\bf 149} 1559--73)
\bibitem{Rudnicki1}Rudnicki {\L} 2011 Shannon entropy as a measure of uncertainty in positions and momenta\JRLR{\bf 32} 393--9
\bibitem{Olendski3}Olendski O 2015 Comparative analysis of electric field influence on the quantum wells with different boundary conditions. I. Energy spectrum, quantum information entropy and polarization\APB{\bf 527} 278--95
\bibitem{Olendski1}Olendski O 2016 Theory of the Robin quantum wall in a linear potential. I. Energy spectrum, polarization and quantum-information measures\APB{\bf 528} 865--81
\bibitem{Olendski4}Olendski O 2018 Quantum information measures of the one-dimensional Robin quantum well\APB{\bf 530} 1700324
\bibitem{Lutz1}Lutz E 2003 Anomalous diffusion and Tsallis statistics in an optical lattice\PRA{\bf 67} 051402
\bibitem{Douglas1}Douglas P, Bergamini S and Renzoni F 2006 Tunable Tsallis distributions in dissipative optical lattices \PRL{\bf 96} 110601
\bibitem{Liu1}Liu B and Goree J 2008 Superdiffusion and non-Gaussian statistics in a driven-dissipative 2D dusty plasma \PRL{\bf 100} 055003
\bibitem{Pickup1}Pickup R M, Cywinski R, Pappas C, Farago B and Fouquet P 2009 Generalized spin-glass relaxation \PRL{\bf 102} 097202
\bibitem{Tang1}Tang Z, Xu Y, Ruan L,van Buren G, Wang F and Xu Z 2009 Spectra and radial flow in relativistic heavy ion collisions with Tsallis statistics in a blast-wave description \PRC{\bf 79} 051901
\bibitem{Shao1}Shao M, Yi L, Tang Z, Chen H, Li C and Xu Z 2010 Examination of the species and beam energy dependence of particle spectra using Tsallis statistics \jpg{\bf 37} 085104
\bibitem{Beck1}Beck C and Schl\"{o}gl F 1997 \emph{Thermodynamics of Chaotic Systems} (Cambridge: Cambridge)
\bibitem{Baez1}Baez J C 2011 R\'{e}nyi entropy and free energy arxiv:1102.2098
\bibitem{Luitz1}Luitz D J, Plat X, Laflorencie N and Alet F 2014 Improving entanglement and thermodynamic R\'{e}nyi entropy measurements in quantum Monte Carlo\PRB{\bf 90} 125105
\bibitem{Mora1}Mora T and Walczak A M 2016 R\'{e}nyi entropy, abundance distribution, and the equivalence of ensembles\PRE{\bf 93} 052418
\bibitem{Helrich1}Helrich C S 2009 \emph{Modern Thermodynamics with Statistical Mechanics} (Berlin: Springer)
\bibitem{Walker1}Walker J 2014 \emph{Fundamentals of Physics} (Hoboken: Wiley)
\bibitem{Bialynicki1}Bia{\l}ynicki-Birula I 2006 Formulation of the uncertainty relations in terms of the R\'{e}nyi entropies\PRA{\bf 74} 052101
\bibitem{Zozor1}Zozor S and Vignat C 2007 On classes of non-Gaussian asymptotic minimizers in entropic uncertainty principles\PA{\bf 375} 499--517
\bibitem{Bialynicki2}Bia{\l}ynicki-Birula I and Mycielski J 1975 Uncertainty relations for information entropy in wave mechanics\CMP{\bf 44} 129--32
\bibitem{Beckner1}Beckner W 1975 Inequalities in Fourier analysis\AnM{\bf 102} 159--82
\bibitem{Beckner2}Beckner W 1975 Inequalities in Fourier analysis on $R^n$\PNAS{\bf 72} 638--41
\bibitem{Campbell1}Campbell L L 1965 A coding theorem and R\'{e}nyi's entropy {\em Inform. Control} {\bf 8} 423--9
\bibitem{Bialas1}Bialas A and Czyz W 2000 Renyi entropies in multiparticle production \emph{Acta Phys. Pol. B} {\bf 31} 2803--17
\bibitem{Bialas2}Bialas A and Czyz W 2000 Event by event analysis and entropy of multiparticle systems\PRD{\bf 61} 074021
\bibitem{Essex1}Essex C, Schultzky C, Franz A and Hoffmann K H 2000 Tsallis and R\'{e}nyi entropies in fractional diffusion
and entropy production\PA{\bf 284} 299--308
\bibitem{Franchini1}Franchini F, Its  A R and Korepin V E 2008 Renyi entropy of the XY spin chain \jpa {\bf 41} 025302
\bibitem{Ansari1}Ansari M H and Nazarov Y V 2015 Exact correspondence between Renyi entropy flows and physical flows\PRB{\bf 91} 174307
\bibitem{Klebanov1}Klebanov I R, Pufu S S, Sachdev S and Safdi B R 2012 R\'{e}nyi entropies for free field theories \emph{J. High Energy Phys.} {\bf 2012} 74
\bibitem{Chen1}Chen B and Zhang J 2013 On short interval expansion of R\'{e}nyi entropy \emph{J. High Energy Phys.} {\bf 2013} 164
\bibitem{Dong1}Dong X 2016 The gravity dual of R\'{e}nyi entropy \emph{Nature Commun.} {\bf 7} 12472
\bibitem{Islam1}Islam R, Ma R, Preiss P M, Tai M E, Lukin A, Rispoli M and Greiner M 2015 Measuring entanglement entropy in a quantum many-body system \emph{Nature (London)} {\bf 528} 77--83
\bibitem{Uffink1}Uffink J B M and Hilgevoord J 1985 Uncertainty principle and uncertainty relations\FP{\bf 15} 925--944
\bibitem{Majernik1}Majern\'{i}k V and Richterek L 1997 Entropic uncertainty relations \EJP {\bf 18} 79-89
\bibitem{Bialynicki3}Bia{\l}ynicki-Birula I 2007 R\'{e}nyi entropy and the uncertainty relations \emph{AIP Conf. Proc.} {\bf 889} 52--61
\bibitem{Bialynicki4}Bia{\l}ynicki-Birula I and Rudnicki {\L} 2011 in: {\em Statistical Complexity: Applications in Electronic Structure}, ed. by Sen K D (Dordrecht: Springer), chap. 1.
\bibitem{Geilikman1}Geilikman M B, Golubeva T V ans Pisarenko V F 1990 Multifractal patterns of seismicity \emph{Earth Planet. Sci. Lett.} {\bf 99} 127--32
\bibitem{Carranza1}Carranza M L, Acosta A and Ricotta C 2007 Analyzing landscape diversity in time: The use of R\'{e}nyi's generalized entropy function \emph{Ecol. Indic.} {\bf 7} 505--10
\bibitem{Drius1}Drius M, Malavasi M, Acosta A T R, Ricotta C and Carranza M L 2013 Boundary-based analysis for the assessment of coastal dune landscape integrity over time \emph{Appl. Geogr.} {\bf 45} 41--8
\bibitem{DeLuca1}De Luca E, Novelli C, Barbato F, Menegoni P, Iannetta M and Nascetti G 2011 Coastal dune systems and disturbance factors: Monitoring and analysis in central Italy \emph{Environ. Monit. Assess.} {\bf 183} 437--50
\bibitem{Rocchini1}Rocchini D, Delucchi L, Bacaro G, Cavallini P, Feilhauer H, Foody G M, He K S, Nagendra H, Porta C, Ricotta C, Schmidtlein S, Spano L D, Wegmann M and Neteler M 2013 Calculating landscape diversity with information-theorybased indices: a GRASS GIS solution \emph{Ecol. Inf.} {\bf 17} 82--9
\bibitem{Jizba2}Jizba P, Kleinert H and Shefaat M 2012 R\'{e}nyi's information transfer between financial time series\PA{\bf 391} 2971--89
\bibitem{Jizba3}Jizba P and Korbel J 2014 Multifractal diffusion entropy analysis: Optimal bin width of probability histogram\PA{\bf 413} 438--58
\bibitem{Rosso1}Rosso O A, Martin M T, Figliola A, Keller K and Plastino A 2006 EEG analysis using wavelet-based information tools \emph{J. Neurosci. Meth.} {\bf 153} 163--82
\bibitem{Costa1}Costa M, Goldberger A L and Peng C-K 2005 Multiscale entropy analysis of biological signals\PRE{\bf 71} 021906
\bibitem{Tozzi1}Tozzi A, Peters J F and \c{C}ankaya M N 2018 The informational entropy endowed in cortical oscillations \emph{Cogn. Neurodyn.} {\bf 12} 501--7
\bibitem{Peters1}A-iyeh E and Peters J 2016 R\'{e}nyi entropy in measuring information levels in Vorono\"{i} tessellation cells with application in digital image analysis \emph{Theor. Appl. Math. Comp. Sci.} {\bf 6} 77--95
\bibitem{Tsallis2} Tsallis C 2009 \emph{Introduction to Nonextensive Statistical Mechanics} (New York: Spinger)
\bibitem{SanchezRuiz1}S\'{a}nchez-Ruiz J 1999 Asymptotic formulae for the quantum R\'{e}nyi entropies of position: Application to the infinite well \JPA{\bf 32} 3419--32
\bibitem{Aptekarev1}Aptekarev A I, Dehesa J S, S\'{a}nchez-Moreno P and Tulyakov  D N 2012 R\'{e}nyi entropy of the infinite well potential in momentum space and Dirichlet-like trigonometric functionals\JMC{\bf 50} 1079--90
\bibitem{Toranzo1}Toranzo I V and Dehesa J S 2016 R\'{e}nyi, Shannon and Tsallis entropies of Rydberg hydrogenic systems \EPL{\bf 113} 48003
\bibitem{Dehesa1}Dehesa J S, Toranzo I V and Puertas-Centeno D 2016 Entropic measures of Rydberg-like harmonic states\IJQC{\bf 117} 48--56
\bibitem{Aptekarev2}Aptekarev A I, Tulyakov D N, Toranzo I V and Dehesa J S 2016 R\'{e}nyi entropies of the highly-excited states of multidimensional harmonic oscillators by use of strong Laguerre asymptotics\EPJB{\bf 89} 85
\bibitem{PuertasCenteno1}Puertas-Centeno D, Toranzo I V and Dehesa J S 2018 R\'{e}nyi entropies for multidimensional hydrogenic systems in position and momentum spaces \JSTAT{\bf 2018} 073203
\bibitem{Abramowitz1}Abramowitz M and Stegun  I A  1964 \textsl{\textit{Handbook of Mathematical Functions }} (New York: Dover)
\bibitem{Saha1}Saha A, Talukdar B and Chatterjee S 2017 On the realization of quantum Fisher information \EJP{\bf 38} 025103
\bibitem{Fisher1}Fisher R A 1925 Theory of statistical estimation {\em Math. Proc. Cambridge Philos. Soc.} {\bf 22} 700--25
\bibitem{Frieden1}Frieden B R 2004 {\em Science from Fisher Information} (Cambridge: Cambridge)
\bibitem{Olendski2}Olendski O 2017 Comment on 'On the realisation of quantum Fisher information' \EJP{\bf 38} 038001
\bibitem{Landau1} Landau L D and Lifshitz E M 1977 {\em Quantum Mechanics (Non-Relativistic Theory)} (New York: Pergamon)
\bibitem{Lebedev1}Lebedev N N 1965 {\em Special Functions and Their Applications} (Prentice-Hall: Englewood Cliffs)
\bibitem{Seba1}\v{S}eba P 1985 Schr\"{o}dinger particle on a half line \LMP{\bf 10} 21--7
\bibitem{Pazma1}Pa\v{z}ma V and Pre\v{s}najder P 1989 On quantum mechanics on a half-line \EJP{\bf 10} 35--7
\bibitem{Bonneau1}Bonneau G, Faraut J and Valent G 2001 Self-adjoint extensions of operators and the teaching of quantum mechanics\AJP{\bf 69} 322--31
\bibitem{Fulop1}F\"{u}l\"{o}p T, Cheon T and Tsutsui I 2002 Classical aspects of quantum walls in one dimension\PRA{\bf 66}  052102
\bibitem{Belchev1}Belchev B and Walton M A 2010 On Robin boundary conditions and the Morse potential in quantum mechanics \jpa{\bf 43} 085301
\bibitem{Georgiou1}Georgiou O, Gligori\'{c} G, Lazarides A, Oliveira D F M, Bodyfelt J D and Goussev A 2012 Influence of boundary conditions on quantum escape \EPL{\bf 100} 20005
\bibitem{Bersons1}Bersons I and Veilande R 2004 Analytical investigation of one-dimensional Rydberg atoms interacting with half-cycle pulses\PRA{\bf 69} 043408
\bibitem{P1atzman1}P1atzman P M and Dykman M I 1999 Quantum computing with electrons floating on liquid helium {\em Science} {\bf 284} 1967--9
\bibitem{Omiste1}Omiste J J, Y\'{a}\~{n}ez R J and Dehesa  J S 2010 Information-theoretic properties of the half-line
Coulomb potential\JMC{\bf 47} 911--28
\bibitem{Arfken1}Arfken G B, Weber H J and Harris F E 2013 {\em Mathematical Methods for Physicists} (Amsterdam: Elsevier)
\bibitem{Messiah1}Messiah A 1961 {\em Quantum Mechanics} vol. 1 (Amsterdam: North-Holland)
\bibitem{Bateman1}Erd\'{e}lyi A 1953 {\em Higher Transcendental Functions} vol. 2 (New York: McGraw-Hill)
\bibitem{Podolsky1}Podolsky B and Pauling L 1929 The momentum distribution in hydrogen-like atoms \PR{\bf 34}
109--16
\bibitem{DePalma1}De Palma G, Trevisan D, Giovannetti V and Ambrosio L 2018 Gaussian optimizers for entropic inequalities in quantum information \JMP {\bf 59} 081101
\end{thebibliography}
\end{document}